# THE DUAL-AXIS CIRCUMSTELLAR ENVIRONMENT OF THE TYPE IIn SUPERNOVA 1997EG

Short title: Circumstellar Environment of SN 1997eg

JENNIFER L. HOFFMAN,[1,2,3] DOUGLAS C. LEONARD,[4] RYAN CHORNOCK,[1]
ALEXEI V. FILIPPENKO,[1] AARON J. BARTH,[5] AND THOMAS MATHESON[6]

[1] Department of Astronomy, University of California, Berkeley, 601 Campbell Hall, Berkeley, CA 94720-3411; alex@astro.berkeley.edu; chornock@astro.berkeley.edu .
[2] NSF Astronomy and Astrophysics Postdoctoral Fellow.
[3] Current address: University of Denver, 2112 E. Wesley Ave., Denver, CO 80208; jennifer.hoffman@du.edu .
[4] Department of Astronomy, San Diego State University, PA-210, 5500 Campanile Drive, San Diego, CA 92182-1221; leonard@sciences.sdsu.edu .
[5] Department of Physics and Astronomy, University of California, 4129 Frederick Reines Hall, Irvine, CA 92697-4575; barth@uci.edu .
[6] National Optical Astronomy Observatory, 950 N. Cherry Ave., Tucson, AZ 85719-4933; matheson@noao.edu .

ABSTRACT

We present multi-epoch spectral and spectropolarimetric observations of the Type IIn supernova (SN) 1997eg that indicate the presence of a flattened disk-like concentration of circumstellar material surrounding nonspherical ejecta, with which the disk is misaligned. The polarization across the broad Hα, Hβ, and He I λ5876 lines of SN 1997eg forms closed loops when viewed in the Stokes *q–u* plane. Such loops occur when the geometrical symmetry of one or both of the Stokes parameters across spectral lines is broken, in this case most likely by occultation of the ejecta by the equatorial circumstellar matter concentration. The polarization of the narrow Balmer lines possesses an intrinsic axis that differs by 12° from that of the elongated ejecta and probably indicates the orientation of the disk-like circumstellar material. The existence of two different axes of symmetry in SN 1997eg suggests that neither rotation of the progenitor nor the influence of a companion star can be the sole mechanism creating a preferred axis within the supernova system. Our model supports the emerging hypothesis that the progenitors of some Type IIn supernovae are luminous blue variable stars, whose pre-supernova mass eruptions form the circumstellar shells that physically characterize the SN IIn subclass. These conclusions, which are independent of interstellar polarization effects, would have been unobservable with only a single epoch of spectropolarimetry.

*Subject headings:* circumstellar matter—polarization—stars: mass loss—supernovae: individual (SN 1997eg)—techniques: polarimetric

## 1. INTRODUCTION

In recent years it has become clear that many if not all supernova (SN) explosions are inherently aspherical processes, whether they arise from thermonuclear (Reinecke, Hillebrandt, & Niemeyer 2002; Gamezo, Khokhlov, & Oran 2005; Kozma 2005) or core-collapse (Burrows et al. 2006; Maeda, Mazzali, & Nomoto 2006; Matt, Frank, & Blackman 2006) scenarios. Observational indications of deviation from spherical symmetry in the geometry of SN ejecta include the elongated debris cloud and complex ring structure of SN 1987A (e.g., Wang & Hu 1994; Wang et al. 2002; Jansen & Jakobsen 2001), asymmetric structures and velocities in SN remnants (e.g., Fesen et al. 2006; Hwang et al. 2004), emission line profiles that suggest bipolar ejecta structures (Chugai et al. 2005; Taubenberger et al. 2006), high-velocity absorption lines in SN spectra (Mazzali et al. 2005), and net polarization of the continuum light (e.g., Wang et al. 1996; Leonard & Filippenko 2001; Wang et al. 2001; Leonard et al. 2006; Pereyra et al. 2006). In addition to continuum polarization, which points to global asymmetries, line polarization features can provide specific clues to the nature of SN ejecta and their surrounding circumstellar material (CSM; Leonard et al. 2000a; Kawabata et al. 2002; Kasen et al. 2003; Wang et al. 2003a,b; Wang et al. 2004; Filippenko & Leonard 2004; Leonard & Filippenko 2005; Leonard et al. 2005; Chornock et al. 2006; Chornock & Filippenko 2008).

Despite indications that the continuum and line polarization characteristics of all types of supernovae (SNe) can change dramatically with time (Cropper et al. 1998; Wang et al. 2003a; Leonard et al. 2006), multi-epoch spectropolarimetry of SNe is still rare. We present here flux spectra at 12 epochs and polarization spectra at 3 epochs for the Type IIn ("narrow-line") SN 1997eg; this array of complementary observations makes it possible to construct a detailed picture of a core-collapse supernova and its surrounding CSM over the first 1.5 years of evolution. We find indications that SN 1997eg is characterized by two distinct axes of symmetry, suggesting that the mechanism creating a preferred axis in the mass-loss phase is either different from the mechanism governing the elongation of the subsequent supernova or else favors varying directions over time.

Type IIn supernovae, first explicitly defined by Schlegel (1990; but see also Filippenko 1989, 1991a,b), are hydrogen-rich (Type II) events whose spectra are characterized by strong, relatively narrow ($v \approx 1000$ km $s^{-1}$; more correctly "intermediate-width") emission lines superposed on much broader bases; the broad P Cygni absorption components seen in normal SNe II are weak or absent in SNe IIn. The spectral continua tend to be quite blue and evolve slowly, while the light curves also tend to show unusually slow declines with time (see the review by Filippenko 1997 for specific examples and references). Cappellaro et al. (1997) estimated that the intrinsic frequency of SNe IIn is 2–5% of all SNe II; however, their sample included only two examples. More recent work with a larger sample by Leaman et al. (2008, in prep.) suggests this frequency may be as high as 15%.

The canonical Type IIn supernova is SN 1988Z (Filippenko 1991a,b; Stathakis & Sadler 1991; Turatto et al. 1993; Chugai & Danziger 1994), but the subclass is quite heterogeneous, with members showing wide variations in strength and shape of emission lines, light-curve behavior, and X-ray and radio luminosity (Filippenko 1997, and references therein; Hoffman 2007, and references therein). Nevertheless, it is generally assumed that the intermediate-width ("narrow") lines in a SN IIn spectrum indicate that the supernova is surrounded by circumstellar material expelled from the progenitor star prior to the SN explosion; this CSM is subsequently excited when the expanding supernova ejecta collide with it. (Photoionization of the CSM by the UV and X-ray photons produced during the initial supernova explosion flash often produces very narrow emission lines in addition to those of intermediate width.) Hence, the characteristics of the CSM provide a link to the mass-loss episodes and stellar winds of the progenitor star in its late stages of evolution. Spectropolarimetry can be a powerful tool in this effort, as it allows us to probe the nature and geometrical structure of the CSM of SNe IIn to an extent impossible with spectroscopy alone (Leonard et al. 2000a; Wang et al. 2001).

SN 1997eg, located along a spiral arm of the SAB(rs)c galaxy NGC 5012 (Fig. 1), was discovered by M. Aoki on 1997 December 4 (Nakano & Aoki 1997; we use UT dates throughout this paper) and spectroscopically classified as Type IIn on 1997 December 20 (Filippenko & Barth 1997). Because NGC 5012 was behind the Sun prior to November 1997, the last null observation was nearly four months before the supernova's discovery: 1997 August 14 (W. Li 2006, private communication) with the 0.76-m Katzman Automatic Imaging Telescope (KAIT) at Lick Observatory (Filippenko et al. 2001; Filippenko 2005). It is therefore difficult to determine the exact age of SN 1997eg at the time of discovery. However, we note that its spectra feature multi-component emission lines of hydrogen and helium similar to those seen in the well observed Type IIn SN 1988Z (Turatto et al. 1993; see § 3.1). Because the optical spectra of SNe IIn show considerable variation, as noted above, this similarity suggests that SN 1997eg and SN 1988Z may be closely related within the heterogeneous SN IIn subclass. The explosion date of SN 1988Z is also not well known, but its extreme luminosity suggests that it was discovered near maximum brightness, probably not more than a few days past core collapse (Stathakis & Sadler 1991; Turatto et al. 1993). Comparing the day 16 spectrum of SN 1997eg with published spectra of SN 1988Z (Stathakis & Sadler 1991; Filippenko 1991a,b), we estimate that SN 1997eg was between 1 and 2 months old at the time of discovery (but see § 4.3 for another estimate).

Tsvetkov & Pavlyuk (2004) obtained *BVRI* photometry of SN 1997eg from 1998 February 21 through March 1 (79–87 d post-discovery), finding that the supernova remained nearly constant in brightness over this period. Using a distance modulus $\mu = 33.03$ mag for NGC 5012 (Tully 1998, assuming $H_0 = 75$ km s$^{-1}$ Mpc$^{-1}$), they calculated that at maximum brightness, SN 1997eg likely had an absolute magnitude $-19 < M_V < -18$. This magnitude estimate places SN 1997eg in the upper range of the luminosity distribution for SNe IIn (Richardson et al. 2002), though it was still significantly fainter than the overluminous, peculiar Type IIn SN 1997cy ($M_V \leq -20.1$ mag; Turatto et al. 2000). Its maximum luminosity is comparable to that of the Type IIn SNe 1998S ($M_V = -18.6$ mag; Barbon et al. 1999; Tully 1998, with $H_0$ as above) and 1995G ($M_V \approx -18.5$ mag; Pastorello et al. 2002).

Filippenko & Barth (1997) found that the spectrum of SN 1997eg consisted of a featureless blue continuum, strong broad emission lines of H and He, and narrow high-excitation emission lines including He II λ4686 and [Fe X] λ6375. They also identified broad, intermediate, and narrow components of the Hα line and noted that the broad component of He I λ5876 was nearly as strong as that of Hα. Salamanca, Terlevich, & Tenorio-Tagle (2002) presented echelle observations of SN 1997eg taken at 198 days post-discovery, in which they detected narrow P

Cygni absorption components superposed on the broad Balmer emission lines. From these observations, they deduced that SN 1997eg is surrounded by dense circumstellar material with a number density $n >\sim 5 \times 10^7$ cm$^{-3}$ and an expansion velocity $v \approx 160$ km s$^{-1}$. Gruendl et al. (2002) conducted similar echelle observations nearly a year later, on 1999 March 2, and found that the narrow P Cygni profile in H$\alpha$ had vanished.

A subset of SNe IIn (most notably SNe 1986J and 1988Z) have become radio-loud objects with peak 6-cm luminosities of ~ $10^{28}$ erg s$^{-1}$ Hz$^{-1}$ (Van Dyk et al. 1996; Pooley et al. 2002). An analysis of the optical spectra of these objects suggests that the radio-loud SNe IIn are surrounded by particularly dense circumstellar material, as inferred from [O III] line intensity ratios and the initial absence of broad H$\alpha$ emission (Filippenko 1997). From the spectrum of SN 1997eg obtained on 1997 December 20, Filippenko & Barth (1997) estimated the CSM density to be ~ $10^8$ cm$^{-3}$ (see § 3.1 for a refined calculation); based on this estimate, they predicted that SN 1997eg would become radio loud. Indeed, SN 1997eg was detected in the radio at a flux density of 0.52 mJy at 3.6 cm (8.44 GHz) on both days 177 and 186 post-discovery (Lacey & Weiler 1998). Again taking the distance modulus $\mu = 33.03$ mag, and assuming the explosion occurred 30 days before discovery, we calculate a radio luminosity of ~ $2.5 \times 10^{26}$ erg s$^{-1}$ Hz$^{-1}$ at approximately 207 days post-explosion. Though this luminosity is several times fainter than that of SN 1988Z at a comparable age (~ $1.2 \times 10^{27}$ erg s$^{-1}$ Hz$^{-1}$ at 3.6 cm from the fits given in Williams et al. 2002), it still places SN 1997eg among the most intrinsically luminous known radio supernovae, confirming the prediction of Filippenko & Barth (1997). From their initial radio data, Lacey & Weiler (1998) calculated that SN 1997eg could eventually reach a 6-cm peak luminosity of ~ $4.9 \times 10^{27}$ erg s$^{-1}$ Hz$^{-1}$, comparable with that of SN 1988Z, but if later radio observations of this SN exist, they have not yet been published.

We discuss our spectral and spectropolarimetric observations of SN 1997eg and associated data reduction procedures in § 2. In § 3 and § 4, we present observational results and detailed analysis; in § 5, we discuss the results and propose an explanatory model. Preliminary results from this dataset were presented by Leonard et al. (2000b).

## 2. OBSERVATIONS AND REDUCTIONS

We observed SN 1997eg with the Low Resolution Imaging Spectrometer (LRIS; Oke et al. 1995) in polarimetry mode (LRIS-P; Cohen 1996) at the Cassegrain focus of the Keck II 10 m telescope during the nights of 1997 December 20, 1998 January 17, and 1998 March 7. We obtained additional flux spectra for SN 1997eg at Keck using LRIS on 1998 March 6 and at the Lick Observatory using the Kast double spectrograph (Miller & Stone 1993) at the Cassegrain focus of the Shane 3 m reflector on eight nights between 1998 January and 1999 March. A journal of observations is given in Table 1. All spectra were taken within 5° of the parallactic angle (Filippenko 1982), except those on days 16, 44, 113, and 402, which were observed at airmasses sufficiently low as to make any relative light losses negligible.

One-dimensional sky-subtracted spectra were extracted optimally (Horne 1986) in the usual manner, generally with a width of ~ 4″ along the slit. When necessary because of the position of the spectrograph slit, background regions were carefully chosen to avoid nearby H II region contamination. Each spectrum was then wavelength and flux calibrated, corrected for continuum atmospheric extinction and telluric absorption bands (Wade & Horne 1988; Matheson et al. 2000b), and rebinned to 2 Å pixel$^{-1}$. We performed the polarimetric analysis according to the methods outlined by Miller, Robinson, & Goodrich (1988) and Cohen et al. (1997). Observations of unpolarized standard stars on all polarimetry nights showed a flat, well behaved polarization response over the observed wavelength range. Similarly, observations of polarized standard stars yielded polarizations which agreed well with catalogued values (Turnshek et al. 1990). Comparison of the position angle (PA) of the polarization of multiple standards observed on the same night revealed discrepancies of ~ 2°, which we adopt as the measurement uncertainty for the polarization PA in this paper. The instrumental polarization of the Keck spectropolarimeter has been found to be <~ 0.1% (Leonard et al. 2005 and references therein).

## 3. RESULTS AND ANALYSIS: TOTAL-FLUX SPECTRA

The total-flux spectra of SN 1997eg are shown in Figure 2. The general spectral characteristics of SN 1997eg closely resemble those of the Type IIn SN 1988Z (Filippenko 1991a,b; Stathakis & Sadler 1991; Turatto et al. 1993). At early times, the continuum flux level increases gradually from red to blue, while very narrow (unresolved), intermediate-width (full width at half-maximum intensity [FWHM] = 2000–3000 km s$^{-1}$), and broad (FWHM ≈ 10,000 km s$^{-1}$) emission lines dominate the spectrum. From day 356 onward, the continuum is quite flat and only a few emission lines, mostly of H I and He I, remain. Line identifications are presented in Table 2 and displayed in Figure 3. Note that the broad and intermediate components of the Balmer lines are sometimes difficult to disentangle, and that the Hβ profile is further complicated by P Cygni absorption, possibly due to an iron blend. We discuss the behavior of individual emission lines in the subsections below.

### 3.1. *Very Narrow Lines*

Very narrow (unresolved) emission lines are prominent in most of the spectra of SN 1997eg. Our spectral resolution puts an upper limit on the width of these lines of FWHM ≈ 200 km s$^{-1}$; Salamanca et al. (2002) measured $v$ = 160 km s$^{-1}$ for the narrow Balmer absorption features in their echelle data. At late times, Balmer and forbidden emission lines most likely arise from a nearby H II region (Gruendl et al. 2002), but careful inspection of the CCD frames, coupled with the changing intensities of the lines over time, confirms that the early-time unresolved lines are associated with the supernova itself. The low velocities of these lines suggest that they arise from a near-stationary CSM distribution outside the supernova ejecta. Like SN 1988Z, SN 1997eg was characterized at early times by "coronal lines" of extremely highly ionized species, including [Fe IV] λ4906; [Fe VII] λλ5159, 5720, 6086; [Fe X] λ6374; [Fe XI] λ7892; and [Fe XIV] λ5303. Though the last of these is only a marginal detection in our spectra, where it is possibly contaminated by [Ca V] λ5309, it was also identified by Salamanca et al. (2002) in spectra near day 200; its presence implies that part of the CSM had early-time electron temperatures of $T_e$ = (0.2–2) × 10$^5$ K if photoionized (Schmidt et al. 2006) and $T_e$ > 2 × 10$^6$ K if collisionally heated (Jordan 1969). The high-excitation narrow lines weaken over time and then disappear from our spectra by day 231.

Narrow low-excitation forbidden emission lines in the spectrum include [O II] λλ7320, 7330; [O III] λλ4363, 4959, 5007; [N II] λ5755; and [Ne III] λ3869. The presence of these lines indicates a cooler component to the CSM (<~ a few × $10^5$ K; Salamanca et al. 2002). As in SN 1995N (Fransson et al. 2002), the absence of [O II] λλ3726, 3729 implies an electron density $n_e > 3 \times 10^4$ cm$^{-3}$, while the ratio $I$([O III] λλ4959 + 5007)/$I$([O III] λ4363) ≈ 1.2–1.3 indicates $n_e$ = (5–40) × $10^6$ cm$^{-3}$ for temperatures in the range $10^5 < T < 10^7$ K (Filippenko & Halpern 1984; Osterbrock 1989). The strong detection of [N II] λ5755 and the non-detection of [N II] λλ6548, 6583 also suggest high density in the CSM; from the signal-to-noise ratio (S/N) in our spectra, we estimate $I$([N II] λλ6548 + 6583)/$I$([N II] λ5755) < 1, implying $n_e > 10^6$ cm$^{-3}$ for $10^5 < T < 10^7$ K (Osterbrock 1989). Potential reddening contributions of $E(B–V)$ = 0.2–0.6 mag (see Appendix) increase these CSM density estimates, which are consistent with those of Salamanca et al. ([5–10] × $10^7$ cm$^{-3}$) and even higher than the density inferred for SN 1988Z for the temperature range adopted by Stathakis & Sadler (1991; [2–4] × $10^6$ cm$^{-3}$).

The compound hydrogen and helium emission lines also show very narrow components. In Figure 4, we present the velocity profiles of the first three hydrogen Balmer lines and two prominent He I lines for SN 1997eg. Because our data are in the form of relative, not absolute, spectra, we have normalized each line profile to a continuum point that we judge to be far from any line emission. In order to emphasize the line shapes, we have varied the vertical scales between panels of this figure. All profiles show narrow lines at zero velocity at early times; these components disappear at varying times from day 55 through day 196.

The behavior of the narrow lines supports a picture in which a slowly moving, high-density region of undisturbed CSM surrounds the expanding ejecta; this region emits narrow Balmer and coronal lines after being ionized by the X-ray/UV pulse from the supernova explosion. In this picture, the disappearance of the narrow lines could be due either to recombination of the CSM or to the CSM being overrun by the SN ejecta. For an H II region of pure hydrogen with $T = 10^5$ K and $10^6 < n_e < 5\times 10^7$ cm$^{-3}$, the expressions derived by Spitzer (1978) lead to an estimated recombination time $t_r$ of 3–168 days, roughly consistent with the timescale on which our observed lines disappear. For $T = 10^7$ K the recombination time is longer (1.3–66 yr), but because $t_r \propto Z^{-2}$, heavier elements are expected to recombine much more quickly than hydrogen. Unfortunately, the time resolution and S/N of our observations are insufficient to allow us to verify whether the narrow lines disappear in the order predicted by the $Z^{-2}$ dependence. Alternatively, the disappearance of the lines might imply that the CSM was close enough to the supernova to be overrun by its ejecta beginning around day 55. We discuss these scenarios and their implications in more detail in § 5.

Instead of being excited by ionizing radiation from the explosion, the narrow lines might be produced by shock heating as a result of ejecta-CSM interaction. We consider this unlikely due to the lines’ very early appearance, but because the explosion date is uncertain, we mention it as another possibility. Higher-resolution spectra at early times would help resolve the question. The presence of narrow hydrogen Balmer, He I, and low-excitation forbidden-line emission at times greater than 231 days post-discovery is most likely due to the contribution of a nearby H II region (Gruendl et al. 2002).

### 3.2. *Intermediate-Width Lines*

Lines of intermediate width (FWHM = 2000–3000 km $s^{-1}$) in the spectrum of SN 1997eg appear only as components of the hydrogen Balmer series. Unlike the narrow coronal lines, these intermediate-width lines persist throughout our series of spectra, which ends at day 463, though they are strongest around day 100. The intermediate-width Balmer line components appear to be skewed toward the blue at early times, but become symmetric after day 92. None of the helium or oxygen lines shows an intermediate-width component.

Figure 4*a* demonstrates that after about 100 days, the intermediate component of each of the first three hydrogen Balmer lines in SN 1997eg develops a double-peaked line profile with a central depression. As this effect could, in principle, be introduced by accidental background subtraction of nearby H II region line emission, we took care to ensure that "sky" regions for the spectral extractions did not include any H II contamination. Examination of the background subtracted from each spectrum shows that the double-peaked profiles are intrinsic to the hydrogen lines and not artifacts of our reduction procedure. These profiles may exist at earlier times but be masked by the narrow lines discussed above. Each intermediate-width Balmer line profile is also complicated by its associated broad component, but because the intrinsic shape of each broad line, especially at later times, is nearly symmetric about its rest wavelength (§ 3.3), its contribution tends to fill in rather than enhance the central depression.

Such a central depression in intermediate-width emission lines was predicted by Chugai & Danziger (1994), who suggested that the lines of intermediate width in SN 1998Z could arise from interaction of the reverse shock with a dense equatorial belt of material superposed on the rarefied wind (their Model B; however, they ultimately preferred a scenario for SN 1998Z in which the shock interacted with dense clumps embedded in the wind). These authors pointed out that if the inclination angle were sufficiently large, emission lines arising from this equatorial region would feature a central depression in the profiles, although they expected H$\alpha$ to be insensitive to this effect. Gerardy et al. (2000) interpreted a similar but much stronger depression in the multi-peak H$\alpha$ profile of SN 1998S as the signature of an equatorial ring of circumstellar material.

Other models for the intermediate component have been proposed. Fransson et al. (2002) attributed the intermediate component of SN 1995N to freely expanding oxygen-rich ejecta material that had not yet been reached by the reverse shock. The lack of clear intermediate-width components in any oxygen lines would seem to argue against this picture for SN 1997eg, though we note that several of these lines are blended with other strong features and are difficult to isolate. Smith et al. (2008) constructed a different model for the Type IIn SN 2006tf in which the intermediate-width lines arose in the cool dense shell outside the reverse shock. In either of these scenarios, the central depressions in the emission lines could be due to self-absorption effects in a physically narrow but optically thick emitting region (Fransson et al. 2005). In this case, however, our spectropolarimetric observations provide additional support for a flattened circumstellar geometry (see § 3.2), so we cautiously favor the equatorial disk-like concentration of material as the origin for the intermediate component in SN 1997eg; we discuss the question further in § 5.

### 3.3. *Broad H and He Lines*

The broad-line spectrum of SN 1997eg consists mainly of He I and hydrogen Balmer lines, several of which are shown in Figure 4. The prominence of the helium lines is an outstanding spectral feature of SN 1997eg at all times: in addition to the narrow components of He II λλ4686, 5411 and He I λλ3820, 3965, 4472, 5017, 5876, 6678, and 7065, broad components of He I λλ3820, 3965, 5017, 5876, 6678, and 7065 are present as well. The lack of He II lines in the broad-line spectrum limits the temperature of the broad-line region to $T$ <~ 20,000K, while the strength of the broad He I λ5876 line relative to the broad components of the Balmer lines suggests an overabundance of He relative to H in the emission region (Osterbrock 1989; Wegner & Swanson 1996; Filippenko & Barth 1997).

The broad components of the hydrogen Balmer and He I lines decrease in velocity extent with time (Fig. 4). All of these broad lines are also asymmetric (peaked blueward of the rest wavelength) at early times, but become symmetric by ~ 200 days post-discovery; when symmetric, the broad He I lines are remarkably triangular. The early non-Lorentzian line profiles and the lack of a smooth blue continuum in the spectrum of SN 1997eg suggest that these broad lines are not formed by multiple scattering of thermal electrons in a circumstellar shell, as suggested by Chugai (2001) for SN 1998S. Instead, the enhanced helium abundance and the decrease in the broad lines' full width near zero intensity (FWZI) over our observation period (from ~ 24,000 km $s^{-1}$ to ~ 14,000 km $s^{-1}$; Fig. 4) suggest a formation region associated with the "cool dense shell" outside the reverse shock, which encounters progressively smaller expansion velocities as it propagates into the ejecta (Chevalier & Fransson 1994; Chugai & Danziger 1994; Chugai et al. 2004). Alternatively, the broad lines may arise in the unshocked ejecta interior to the reverse shock, as proposed by Smith et al. (2008) for SN 2006tf; in either case the fastest velocities observed correspond closely to the location of the reverse shock. Using the Hubble-type expansion law $v = r/t$ and assuming SN 1997eg exploded 30–60 days prior to discovery (§ 1), we estimate that the radius of the reverse shock was (3–6) $\times 10^{15}$ cm at day 16 and (2–4) $\times 10^{16}$ cm at day 402.

Fransson et al. (2005) noted that if the observed broad-line profiles are peaked instead of boxy, as in this case, the hypothesis that the broad lines arise from the cool dense shell may imply that the circumstellar medium with which the ejecta interact is asymmetric. This finding is consistent with our tentative conclusions in the previous section that the intermediate-width lines are associated with an equatorial concentration of circumstellar material. In this picture, the overabundance of helium in the broad-line spectrum and its absence in the intermediate-width spectrum imply that the progenitor star underwent an episode of mass loss characterized by a slow, dense, equatorially concentrated, hydrogen-rich wind that formed the circumstellar medium, leaving behind a helium-rich atmosphere as the eventual source of the supernova ejecta. This interpretation contrasts with that of SNe 2006jc (Foley et al. 2007) and 1999cq (Matheson et al. 2000a), two other supernovae whose spectra showed prominent helium emission lines. In these latter two cases, the helium lines were of intermediate width and hydrogen was weak or absent, suggesting a helium-rich CSM formed by mass loss from a progenitor that had already lost most or all of its hydrogen.

The broad, blended [Ca II] near-infrared triplet shows a gradual decrease in FWZI throughout our observations similar to those of the H and He lines; however, it disappears between days 225 and 356. The broad feature in the range 5100–5400 Å behaves very differently with time than do the other broad lines; its width remains constant over our observations, and its strength with respect to the continuum is greatest in the middle epochs. This feature is likely due not to a single emission line but to the superposition of numerous Fe II lines, as in SN 2006jc (Foley et al. 2007) and SN 1988Z (Turatto et al. 1993).

## 4. RESULTS AND ANALYSIS: SPECTROPOLARIMETRY

### 4.1. *General Polarization Features*

Figure 5 displays the polarization spectra of SN 1997eg at our three epochs of observation. Because the magnitude of polarization varies both with time and across spectral lines, a significant component of the observed polarization must be intrinsic to the object and not due to polarization by interstellar dust in the Milky Way or in NGC 5012. We recognize, however, that the contribution of interstellar polarization (ISP) to these spectra is likely not negligible. Accordingly, we have tried several techniques to quantify this contribution, but have not found a definitive solution. Due to the complexity of this case, we concentrate here mainly on the polarimetric analysis that can be conducted without knowing the ISP. In § 4.3 and the Appendix, we constrain potential ISP contributions and address their implications for further interpretation of these polarization data.

In order to analyze the continuum polarization, we defined three spectral regions free from strong line emission and line polarization effects at any epoch: 4500–4600 Å ("blue"), 5400–5500 Å ("green"), and 6100–6200 Å ("red"). For each continuum region at each epoch, we calculated the error-weighted mean percent polarization and position angle (PA); these values are given in Table 3. (We used the measured polarization uncertainties for the weighted means, but in the following analysis we take $\sigma_p = 0.1\%$ and $\sigma_\theta = 2°$ for all observations. These values, which are ~10 times larger than the internal uncertainties, represent systematic errors due to instrumental polarization; see § 2.) From the table, we can see that the continuum polarization of SN 1997eg is approximately constant with wavelength in each observation, implying that wavelength-independent electron scattering is the dominant source of polarization at these epochs. The position angle of polarization is also approximately constant with wavelength, suggesting that a single scattering region polarizes all the continuum light. Both the percent polarization and the position angle of the continuum appear to decrease roughly linearly with time, though we caution that correction for ISP (see § 4.3) may change this apparent behavior.

Figure 6 shows the wavelength-averaged continuum polarization measurements of SN 1997eg in the *q–u* plane. The change in polarization of the continuum over time causes these continuum samples to vary nearly along a straight line in this plane. The simplest way to explain this behavior is to hypothesize that the change in polarization with time is only a change in polarization magnitude, while the PA remains constant. This scenario arises naturally in the case of continuum light interacting with an expanding scattering region; the *q–u* line formed by these points corresponds to the orientation of this scattering region in the plane of the sky, while the change in polarization corresponds either to a change in the optical depth of the expanding scattering region or a change in its illumination. (Note that although the *observed* polarization decreases with time, we cannot know without correcting for interstellar polarization whether this change corresponds to a decrease or an increase with time of the *intrinsic* polarization of the supernova; we address this issue in § 4.3.) A least-squares fit to these points, weighted by the *q* and *u* errors of the mean polarization calculations above, yields a position angle for this line of either 35° or 125°. (Recall that position angles in *q*–*u* space are half-angles, so this 90° difference translates to a 180° difference on the sky. The degeneracy arises because without an ISP estimate, we can solve only for the slope of the line, not its direction with respect to the origin of the *q*–*u* plane.) We interpret this PA as the axis of the continuum scattering region. If the continuum light arises from ellipsoidal SN ejecta, as in the standard picture of SN polarization (e.g., Höflich 1991), then these ejecta must be oriented with the long axis either at 125°–305° or at 35°–215° in the plane of the sky (perpendicular to the polarization PA, which traces the polar axis of the ellipsoidal distribution).

## 4.2. *Polarized Line Features*

Because our flux spectra are not absolute, we cannot directly compare the magnitude of the polarized flux between epochs (Fig. 5). We can, however, analyze the spectral features and compare the spectra in velocity space with each other and with the total flux spectra. Figure 5 shows that the polarized flux spectrum of SN 1997eg displays three strong line features: Hα, Hβ, and He I λ5876. Hα and Hβ each possess broad and intermediate components in the polarized flux, while He I λ5876 is present only as a broad line. For clarity in Figure 5, we have binned the polarized flux spectra to 10 Å, thus obscuring any potential narrow-line signatures in the scattered light. In fact, the narrow Hα and Hβ lines do appear to have polarized flux behavior distinct from their broad and intermediate-width components, as we discuss in § 4.2.2 below.

### 4.2.1. Polarized Broad Lines and Loops in *q*–*u* Space

It is apparent from Figure 5 that the line profiles of the broad Hα and Hβ lines differ between the total-flux spectra and the polarized flux spectra; the polarized hydrogen lines have broader and more asymmetric shapes than do the unpolarized lines. (Because the flux spectra in the top panels of Figure 5 represent *total* flux, they each include a small contribution from polarized light. However, this contribution is small enough at each epoch—less than 3%, as can be seen from the magnitude of the polarization in the second panel of each plot—that the total-flux spectrum is essentially the same as the direct-light spectrum.) In Figure 7 we display this phenomenon in more detail, contrasting the behavior of the broad hydrogen lines in flux and polarized flux with that of the broad He I λ5876 line. For ease of comparison in this figure, we have aligned the flux and polarized flux continua in each frame. The polarized Hα and Hβ lines have enhanced broad blue wings compared with their unpolarized counterparts, but no corresponding red features exist. By contrast, the helium line profiles appear the same in polarized and unpolarized light; in the first epoch alone, the polarized line is blueshifted with respect to the line in direct light by approximately 2000 km $s^{-1}$. We note that the breadth in velocity space of these polarized lines is independent of interstellar polarization effects.

The similarity between the polarized and unpolarized line profiles of He I λ5876 implies that this broad line is polarized via scattering in a region whose geometry is similar to that of the region where it was emitted. This is the expected result if the broad lines both arise from ellipsoidal ejecta and become polarized by electron scattering there, as in the standard picture of supernova polarization (e.g., Höflich 1991). However, the fact that the broad and intermediate-width hydrogen lines have different profiles in polarized light than in unpolarized light implies that these lines are polarized by a different mechanism (or combination of mechanisms) than the He I λ5876 line. The presence of the enhanced blue wing of the hydrogen lines suggests that at least part of their polarization arises from scattering in a region that is approaching at a higher velocity than the line-emitting region. If we postulate a flattened, possibly disk-like circumstellar scattering region external to the ejecta, then the lack of an enhanced red wing in the polarized flux profiles of Hα and Hβ may indicate that we are viewing this scattering region nearly edge-on and that its receding half is blocked from our view either by the ejecta or by the approaching side of the scattering region itself. Because the polarized line profiles of He I λ5876 show no enhanced blue wing, we can assume that these wavelengths do not experience significant scattering within the CSM. This supports our model of a hydrogen-rich circumstellar medium (§ 3.3) and suggests that this CSM polarizes via resonance-line scattering rather than electron scattering. The CSM must therefore include both a near-stationary region that gives rise to the narrow coronal lines (§ 3.1) and a region with velocities up to a magnitude of 10,000 km $s^{-1}$ (seen in the enhanced blue Balmer line wings; Fig. 7). Such a situation may arise as the double-shock structure of the supernova makes its way through the CSM; we explore this model further in § 5.

We can gain further insight into the characteristics of this scattering region by plotting the hydrogen and helium lines in Stokes *q*–*u* space, a method that allows us to view all polarimetric information about each line in one plot. In Figure 8, the Hα, Hβ, and He I λ5876 lines are shown as heavy lines and binned to a resolution of 50 Å to emphasize that each traces a nearly closed loop in the *q*–*u* plane. A simple depolarization or dilution effect would appear as a straight line segment in this plane, as neither causes a change in position angle (e.g., Vink et al. 2002). However, when the variation in one Stokes parameter is symmetric about the emission-line center but variation in the other is not, the net effect is a change in polarization *and* a position angle rotation across the line; these combined effects create a loop-like shape in the *q*–*u* plane (McLean 1979) such as those seen in Figure 8.

This signature, which is independent of interstellar polarization effects, often occurs in Herbig Ae stars, where it is commonly interpreted as the signature of a rotating disk; rotation breaks the symmetry in Stokes *u* across the line center (Vink et al. 2002). Loop-like *q*–*u* line shapes are also seen in Seyfert 1 galaxies, where a similar scenario is postulated: light traveling from the redshifted side of the rotating disk to a scattering element forms a different angle to the observer than does light arising from the blueshifted side (Smith et al. 2005; their polarized lines display loop behavior when converted from [*p*, θ] to [*q, u*] and plotted as in Figure 8). In both of these cases, however, the *q*–*u* loops occur over a much smaller velocity range than do the ones in Figure 8. Believing rotational velocities of ~ 18,000 km $s^{-1}$ (values seen in the polarized blue wing of Hα) to be infeasible in the case of SN 1997eg, we turned to another model to explain the loop behavior.

Loops in *q*–*u* space characterize spectral features in a variety of supernovae. Cropper et al. (1988) observed *q*–*u* loops across both the Hα and Ca II λ8600 emission lines of the Type II SN 1987A. More recently, Wang et al. (2003a) and Kasen et al. (2003) found a similar effect in the highly blueshifted Si II and Ca II absorption lines of the Type Ia SN 2001el, while Maund et al. (2007a,b) observed *q*–*u* loops across the He I λ5876 line in the Type Ib/c SN 2005bf and the broad Ca II near-infrared triplet in the Type IIb SN 2001ig. Such loops are usually attributed to an unspecified wavelength-dependent asphericity of the SN ejecta. Kasen et al. (2003), however, constructed detailed geometrical models for SN 2001el and favored a scenario in which a high-velocity clump of material both creates the flux absorption feature and partially obscures the polarized SN atmosphere. They found that this configuration created a geometrical asymmetry that produced offset peaks in the *q* and *u* polarization spectra, leading to a loop-like shape in *q*–*u* space. Further detailed modeling carried out by these authors suggested that such loops arise as a general feature of dual-axis systems, or those in which the emission and absorption/scattering regions are elongated along different spatial directions.

Such a dual-axis model is the simplest way to explain the spectropolarimetric features of SN 1997eg. Because both the double-peaked intermediate-width hydrogen line profiles and the enhanced blue wings of the hydrogen broad lines in the polarized flux suggest the existence of a flattened disk-like or ring-like distribution of circumstellar material around SN 1997eg, we favor a model similar to the toroidal shell discussed by Kasen et al. (2003) rather than the clump scenario they prefer for SN 2001el. In this picture, the continuum polarization arises from electron scattering in the ellipsoidal SN ejecta, while the broad-line polarization of Hα, Hβ, and He I λ5876 arises from obscuration of portions of the ejecta by a toroidal or disk-like CSM whose axis of symmetry is not aligned with that of the ejecta. This misalignment between the two axes in the system breaks the axial symmetry of the ellipsoidal ejecta and creates the offset between Stokes parameters that causes the *q*–*u* loops in Figure 8. That the loops are generally elongated in the same direction as the continuum polarization axis supports this interpretation; if no obscuring CSM existed, these loops would collapse to line segments having the position angle of the ejecta.

#### 4.2.2. Polarized Very Narrow Lines

As the broad and narrow hydrogen lines behave differently in the total-flux spectra of SN 1997eg (§ 3.1), it is not surprising that they also show different behavior in the *q*–*u* plane. The diamond symbols in Figure 8 represent our data (in bins of width 5 Å) for velocities in each line less than 550 km $s^{-1}$. We stress that we have not attempted to isolate the narrow lines from the other components, so the polarization plotted here contains contributions from the continuum, the broad lines, and the intermediate-width lines as well as from the narrow lines. However, plotting the change in polarization across this small wavelength range allows us to identify the axis along which the polarization trends as the contribution from the narrow line varies. This axis should correspond to the physical axis of the narrow-line scattering region (Quirrenbach et al. 1997; Harries et al. 1999). At each epoch, the narrow Hα line traces a linear path in *q*–*u* space whose direction is quite distinct from the direction of elongation of the broad-line loops; the few points defining the Hβ narrow line seem to follow a similar path. In contrast, the points corresponding to the narrow He I λ5876 line fall along a different path, more vertical and less obviously linear.

Figure 9 plots narrow-line points ($|v| < 550$ km $s^{-1}$) from the two hydrogen lines at all epochs together; this figure makes it clear that these narrow lines lie along a different axis in *q*–*u* space than the one defined by the continuum polarization. Least-squares fits to the Hα and Hβ points, weighted by their internal uncertainties, yield position angles of 24° (114°) and 22° (112°), respectively; see § 4.1 for an explanation of the degeneracy. These very similar slopes suggest that the two lines are polarized by the same mechanism in the same geometrical region. The slight separation in *u* between the Hα and Hβ points is likely due to the underlying continuum polarization, which we have not removed and which varies by a few tenths of a percent over the optical spectrum (see Fig. 6 and Table 3). Nevertheless, the fact that these points lie along an axis in *q*–*u* space distinct from that of the continuum provides us with an ISP constraint; it suggests that the narrow Balmer lines are polarized at an intrinsic position angle of approximately 23° (113°) that remains constant with wavelength (for $|v| < 550$ km s$^{-1}$), even as the magnitude of polarization changes. We make use of this constraint in § 4.3.

We note that we cannot rule out the presence of loop-like variations across the narrow lines, as these would be unresolved in our data. (Reduced $\chi^2$ values are near 1 for each linear fit in Figure 9, but this does not exclude loops whose widths are smaller than our error bars.) Even so, it is evident in the *q*–*u* plane that the polarization of the narrow hydrogen lines does not arise from the same scattering region that produces the broad-line polarization. If these narrow lines originate in unshocked, nearly stationary circumstellar material, then they may be polarized by resonance-line scattering in the same flattened CSM distribution postulated above, in which case the 23° (113°) position angle may be interpreted as the axis of that disk-like distribution.

The He I λ5876 emission-line polarization is represented in Figure 9 by the dotted line, which is the result of a least-squares error-weighted fit to the points from the last row of Figure 8. These narrow-line helium points clearly do not lie along the *q*–*u* axis defined by the narrow hydrogen lines, but instead seem to align roughly parallel to the continuum axis in the *q*–*u* plane. Again, we cannot tell whether a loop exists in these narrow-line data. However, the behavior we observe can easily be explained by a scenario in which the narrow helium lines are intrinsically unpolarized, so that the position of their representative points in the *q*–*u* plane is determined by the underlying continuum. That the narrow helium line polarization tends to vary with time in the same way as the continuum polarization (Figs. 6, 8) supports this interpretation. This scenario implies that photons of He I λ5876 scatter very seldom within the CSM (many scatters would produce the *q*–*u* axis of 23° [113°] seen in the narrow hydrogen lines), again consistent with the hypothesis (§ 3.3) that the CSM around SN 1997eg is composed mainly of hydrogen.

#### 4.2.3. Polarized Intermediate-Width Lines and Line Blends

The polarization behavior of the intermediate-width Hα, Hβ, and He I λ5876 lines (not shown in Fig. 8) is difficult to characterize; there are indications of both loop-like and linear features in different lines at different times, with no clear pattern. We suspect that superposition of the broad-line and narrow-line effects in this velocity region complicate efforts to isolate any intrinsic intermediate-width behavior.

The polarization behavior of the broad iron-blend feature at 5100–5400 Å is different from any of the lines so far discussed, which is not surprising considering that this feature is a blend of many different emission lines. The cluster of points in the *q*–*u* plane corresponding to this feature is centered on the continuum line at each epoch, but shows a pronounced spread perpendicular to the continuum line and larger than the error bars on the points (Fig. 10). These points do not form a loop as a function of wavelength like the broad H and He lines, nor do they follow the line traced by the narrow Balmer lines. The behavior of this feature is most likely a combination of that of the narrow [Fe VII] λ5159/[Fe II] λ5160 line, which probably arises from the CSM, and several broad iron lines, especially of Fe II, that are associated with the SN ejecta (Foley et al. 2007). We expect the forbidden emission from the CSM to be intrinsically unpolarized, but the polarization of the underlying continuum would cause this feature to follow the continuum line in *q–u* space. The spread perpendicular to the continuum axis is thus probably due to the broad Fe blend; this behavior suggests that the iron in the ejecta of SN 1997eg is concentrated near the equator of the ejecta instead of along its major axis, which is defined by the continuum line. Evidence for such stratification in composition has now been found in several core-collapse SNe, including SN 1987A (Wang et al. 2002) and SN 2002ap (Wang et al. 2003b).

### 4.3. Interstellar Contributions and Limits on Intrinsic Polarization

Up to this point, the results we have presented from our spectropolarimetric measurements of SN 1997eg have been independent of any interstellar polarization (ISP) effects. In fact, it has proven quite difficult to establish a well-defined value of ISP for this object. Here we briefly summarize the results of our analysis of the potential ISP contributions to the polarization of SN 1997eg, deferring a more detailed discussion to the Appendix.

Our main assumption in constraining the ISP is that the variations with time we observe in the continuum polarization of SN 1997eg are changes only in the magnitude of polarization and not in position angle (§ 4.1). This assumption implies that the point in *q–u* space representing the ISP lies along the dashed line in Figure 6, which describes the variation of the continuum polarization. In this figure, we have marked three representative ISP estimates (Cases A, B, and C) whose justifications and implications we describe more fully in the Appendix. Our preferred estimate, Case A, corresponds to the intersection in *q–u* space of the continuum line with a line representing the polarization of the Hα and Hβ narrow emission features (Fig. 9; § 4.2). This choice implies both that the intrinsic continuum polarization of SN 1997eg increased with time while retaining a constant PA (as did that of SN 1987A; Jeffery 1991) and that the narrow Balmer emission lines share a common polarization axis that is also constant with time but distinct from that of the continuum. We find $p_{\max,\,A} = 2.9\%$ and $\theta_A = 111°$ for the ISP in this case.

Table 4 compiles the intrinsic continuum polarization values of SN 1997eg after correction for each of our three ISP estimates (see Appendix for details). In Figure 11, we plot these intrinsic values as a function of time and extrapolate the resulting linear variations back to pre-discovery epochs. The shaded boxes in Figure 11 delineate the extrapolated intrinsic polarization for SN 1997eg at the time of explosion ($t_{\rm exp}$) in each case, taking into account the uncertainties in our fitted lines. For Cases B and C (and potential choices between them and beyond Case C), we have assumed $t_{\rm exp}$ = 30–60 d pre-discovery, as discussed in § 1. In the illustrative models of Höflich (1991), the lower limit of $p \approx 3\%$ for choices in this range would imply an axis ratio $E$ <~ 0.5 for an oblate spheroidal scattering atmosphere viewed edge-on. Explosion dates much earlier than 60 days pre-discovery lead to extrapolated intrinsic continuum polarization values significantly larger than any yet observed in a supernova (Wang et al. 2001; Leonard &

Filippenko 2005; Gorosabel et al. 2006); this suggests that our estimate of $t_{\rm exp}$ based on spectral comparison with SN 1988Z (§ 1) is reasonable.

Figure 11 also shows that removing our preferred ISP choice, Case A, from the observed data leads immediately to an independent estimate of the explosion date for SN 1997eg as the epoch at which the intrinsic polarization decreases to zero (see Appendix for details). This analysis yields $t_{\rm exp} = 9^{+45}_{-9}$ days pre-discovery, which (given the large uncertainties discussed in the Appendix) is again consistent with our previous estimate.

All of our ISP estimates lead to large values of intrinsic polarization (> 2%) for at least some of the narrow hydrogen Balmer line points shown in Figures 8 and 9. In Case A, we find that the narrow Hα and Hβ lines are intrinsically polarized between 0.4% and 2.5% at a constant PA of 23° (113°). Because we have not attempted to remove the underlying polarization contributions from the continuum or from the broad and intermediate-width lines, the actual narrow-line polarization may be smaller. However, the fact that the narrow Balmer line polarization varies with time in a very different way from the continuum or the other line components provides a strong indication that these lines are not intrinsically unpolarized, as is sometimes assumed in spectropolarimetric analyses. This is a good argument for the utility of multiple epochs of spectropolarimetry; with only one polarimetric observation, it would have been much more difficult to reach this conclusion.

## 5. DISCUSSION

We seek a model for the structure of the ejecta and circumstellar material of SN 1997eg that explains the following observations:

(1) very narrow, high-excitation emission lines and narrow lines of hydrogen and helium that fade by day 200 (§ 3.1);

(2) intermediate-width hydrogen lines that show double-peaked profiles between days 100 and 250 (§ 3.2);

(3) a polarized continuum whose magnitude changes linearly with time, at an intrinsic PA of ~ 35° or 125° (§ 4.1);

(4) polarized broad lines of Hα and Hβ that show an enhanced blue wing compared with their unpolarized counterparts (§ 4.2.1);

(5) polarized broad line of He I λ5876 that mimics the shape of its unpolarized counterpart and whose profile is blueshifted by 2000 km s$^{-1}$ on day 16 (§ 4.2.1);

(6) polarized broad lines of Hα, Hβ, and He I λ5876 that show well-defined loops in *q–u* space (§ 4.2.1);

(7) polarized narrow lines of Hα and Hβ with an intrinsic PA of ~ 23° or 113° (§ 4.2.2); and

(8) polarized narrow lines of He I λ5876 with an intrinsic PA agreeing closely with that of the continuum (§ 4.2.2).

As discussed in the sections above, observation (1) suggests the presence of a distribution of CSM that is ionized by high-energy emission from the supernova and later recombines or is overrun by the ejecta. Observations (2) and (4) suggest that a dense equatorial region, viewed nearly edge-on, both gives rise to intermediate-width hydrogen emission lines and scatters photons contributing to the broad hydrogen emission lines. However, observations (4) and (6) tell us that the regions where the broad Balmer lines are emitted and scattered cannot have the same geometrical structure. Conditions (3), (7), and (8) taken together also imply the presence of at least two distinct scattering regions in the system. Observation (5) is the most difficult to explain, as it implies a difference in the behavior between the broad hydrogen and helium lines, in apparent contradiction to (6).

### 5.1. Dual-Axis Model

Given these conditions, we propose the following basic picture (Fig. 12), for the moment assuming Case A for the ISP (§ 4.3; Appendix) so that all data points translate into the first quadrant of the *q*–*u* plane. We interpret the 35° position angle as the signature of the aspherical SN ejecta, which we depict here as an ellipsoid whose long axis is oriented at 125° on the sky (§ 4.1). These ejecta are interacting with circumstellar material composed mainly of hydrogen, with which they are misaligned; that is, the CSM does not share an axis of symmetry with the ejecta. In particular, we envision the CSM as having an axis oriented at 23° on the sky and an equatorial disk-like region of enhanced density ($n_e \gtrsim 10^7$ cm$^{-3}$; § 3.1) that is therefore elongated at 113°. The contribution of interstellar polarization changes only the absolute angle measures given here and in Figure 12; the difference of approximately 12° between the axes of the two geometrical structures is intrinsic to our data and is independent of ISP. We also suppose that the inclination of the disk region is nearly edge-on to our line of sight.

In this picture, continuum light both originates and scatters via electron scattering within the ejecta, whose expansion produces a change in polarization magnitude but not PA over time. The interaction of the He-rich ejecta with the H-rich CSM gives rise to intermediate-width Balmer emission lines that show the double-peaked signature of the CSM's dense disk-like equatorial region, while the reverse shock propagating into the ejecta produces broad emission lines of both hydrogen and helium. Light from these broad lines is also polarized by electron scattering within the ejecta, but obscuration of portions of the ejecta by the misaligned disk produces the observed loops in *q*–*u* space across the broad lines. For this obscuration (absorption or scattering) to occur, part of the disk-shaped CSM must already have been accelerated by the forward shock to velocities comparable to those of the broad-line photons. We see only the approaching side of the disk, where the broad blue Balmer-line wings acquire an enhanced polarization due to resonance scattering. Narrow lines of hydrogen, helium, and iron as well as narrow coronal lines arise from an ionized but nearly stationary ($v \lesssim 200$ km s$^{-1}$; § 3.1) region of the disk outside the shocked region. The narrow Balmer lines experience a resonance-scattering opacity in the disk sufficient to polarize them along its axis, but because the CSM is not helium-rich or iron-rich, the narrow He and Fe lines escape without being highly polarized. Similarly, the low helium abundance in the CSM gives rise to a broad He I λ5876 emission-line profile that is the same in total flux and polarized flux, as well as producing *q*–*u* plane shapes that are much less loop-like than those of the broad Balmer lines. Finally, the ejecta show a stratification in composition such that the equatorial regions (away from the ellipsoidal major axis) are iron-rich; the Fe II lines arising here are polarized perpendicular to the continuum light, creating the observed spread of the 5100–5400 Å iron blend in *q*–*u* space.

We have assumed that the polarization in SN 1997eg is caused primarily by electron scattering; at the relatively early times of our polarimetric observations, the SN environment should still be quite warm and largely ionized. Within the circumstellar material, however, the polarization cannot be due to electron scattering; if it were, all wavelengths would be affected in the same way, and we would expect the hydrogen and helium narrow lines to show the same behavior in the *q*–*u* plane. In this case one would also expect the entire continuum spectrum to form a *q*–*u* loop at each epoch, which it does not. Because only the narrow hydrogen lines appear to scatter in the disk-like CSM, we have postulated resonance scattering as the polarizing mechanism. In normal supernova atmospheres, line scattering is assumed to be depolarizing because the timescale for redistribution of level populations is short compared to the timescale for line absorption and re-emission, so the polarizing effects of the resonance scattering are washed out (Jeffery 1991; Höflich 1996). Our line polarization results suggest that although the density in the CSM around SN 1997eg is relatively high (§ 3.1), collisions may occur infrequently enough to reveal the polarizing nature of the resonance scattering process.

Dust formation in the ejecta is thought to give rise to asymmetric and blueshifted Hα line profiles in some SNe II, but this occurs at later times when the temperature of the ejecta has dropped below a few thousand K, normally more than 200 days post-explosion (e.g., Elmhamdi et al. 2003; Wooden et al. 2003; Pozzo et al. 2004; Sugerman et al. 2006). Our spectra show no pronounced asymmetric profiles in the hydrogen emission lines through day 463 (Fig. 4*a*), so we assume that dust scattering is not a significant contributor to the polarization we observe at day 93 and earlier. We also note that although a dusty disk could produce the obscuration that gives rise to the broad-line *q*–*u* loops, it would not also give rise to the narrow-line polarization effects we observe; postulating such a dusty structure would thus amount to introducing a third component into our picture of the system.

This simple model has several potential shortcomings. First, because it requires the ejecta to be partly obscured by the disk-like CSM, it would seem to predict the existence of H and He absorption features in the total-flux spectrum. Due to the misalignment of the disk with the ejecta, these features should occur at two different velocities within each broad-line profile; if the optical depth of the ejecta were sufficiently low, two features would be observed, symmetric about the line center. The strength and breadth of these features would depend on the opacity and physical thickness of the disk, so it is difficult to assess whether we should be able to detect them in our spectra.

Second, the model suggests that the intermediate-width Balmer lines should scatter in the disk before reaching us, which would give them a polarization signature similar to that of the narrow lines. In contrast, we observe no consistent pattern in the polarization of these intermediate lines. As mentioned in § 4.2 above, we attribute this discrepancy to the fact that the light making up the intermediate lines may include contributions from both the broad and narrow components. Models such as those of Chugai & Danziger (1994) and Chugai et al. (2004) posit a heterogeneous “mixing layer” behind the forward shock, created by Rayleigh-Taylor instabilities and characterized by dense knots and filaments within more rarefied gas. Such a clumpy and dynamic environment may well give rise to polarization signatures that lack well-defined structure and vary among different lines and over time.

Third, nothing in this model specifically predicts the second part of observation (6), the blueshifted broad He I λ5876 polarized profile in our earliest observation. We note that it is possible that the Hα and Hβ broad-line components share this blueshift, though the noise in the polarized flux spectra make this difficult to establish with certainty. (However, the intermediate-width polarized Hα line is clearly centered on the rest wavelength in agreement with its total-flux profile.) If the blueshift holds for all three broad lines, perhaps a transient asymmetry in the ejecta expansion velocity produced a shell of ejecta material that on day 16 was moving toward us with abnormally high speed (but with the same geometrical shape as the rest of the ejecta). This scenario is extremely speculative, but it illustrates the usefulness of broad time coverage in spectropolarimetry as well as spectroscopy of supernovae. Tran et al. (1997) found a similar but redshifted effect in the polarized spectrum of the Type IIb SN 1993J, which may hint at a clumpy structure for the He-rich ejecta in both objects. More detailed modeling of the type of scenario presented here would help address such difficulties and would clearly be useful in interpreting such complex data in the future.

As noted above, the difference between the position angles of the two proposed geometrical structures is 35° – 23° = 12°. If we imagine the ejecta to be ellipsoidal in shape with an axis ratio $b/a = 0.5$ (as in the models of Höflich [1991]) and we view the ellipsoid edge-on and projected onto a plane such that its major axis lies along $x$ and its minor axis along $z$ (Fig. 13), then the disk-like CSM blocks the polarization of the underlying ejecta at 12° and 192° from the $x$-axis. The slope of the tangent line to the projected ellipse at both these locations is –1.18, which corresponds to a polarization position angle of –50°. Light polarized at this angle (plus or minus an amount depending on the geometrical thickness of the disk) is blocked from our view by the opaque disk. Recalling that $\theta = 0°$ when all the polarization is in $+q$ and $\theta = -45°$ when all the polarization is in $+u$, we can tell that a disk at this angle will block out mostly light polarized in the $+u$ direction. This results in a net negative $u$ polarization, because the $u$ vectors would otherwise cancel completely. Narrow lines arising in the disk, by contrast, show a net positive polarization in $u$ due to the disk's 12° orientation.

In order to check this interpretation, we would need to rotate our observed polarization vectors into the frame depicted in Figure 13, which requires estimating both the ISP and the orientation of the SN 1997eg system on the sky. For example, if we subtract the Case B ISP estimate from § 4.3 from our data and rotate them by ~ 45° (one counter-clockwise quarter-turn in $q$–$u$ space), the loop shapes in Figure 8 transform so that their long axes lie roughly along the positive $q$ axis, with the continuum endpoints in the second quadrant (positive $q$, negative $u$) and the narrow emission-line points in the first quadrant ($q$ and $u$ both positive). This would imply that the long axis of the ejecta was oriented at –45° on the sky and the disk was oriented at –57°. We stress that this is a very ill-constrained exercise given the available information, but with more extensive data on a future supernova, such a procedure might yield quite good estimates of its spatial orientation.

The close alignment between the polar axis of the ejecta and the dense CSM disk suggests that the ejecta may eventually be decelerated by interaction with the disk material. This deceleration would tend to circularize the ejecta over time, causing the $q$–$u$ loops across the broad emission lines to become lines with the position angle of the disk. Such deceleration apparently did not occur in SN 1997eg in its first 463 days (see § 5.2), but if loop signatures like those analyzed here appear in future supernovae, a detailed sequence of polarimetric observations may allow us to quantify the deceleration of the ejecta and put further limits on the structure of the interaction region.

### 5.2 Implications for the CSM and Progenitor

We now further investigate the nature of the equatorially flattened circumstellar matter distribution exterior to the shocked region in SN 1997eg. Representing this disk-like configuration as a sector of a spherical shell with inner radius $R_{\rm disk}$, equatorial opening half-angle α, density ρ, and fractional radial thickness $f = \Delta R/R_{\rm disk}$ allows us to calculate the circumstellar mass in the disk via

$$\frac{M_{\rm disk}}{M_\odot} \approx \frac{\alpha\rho}{1.416\,{\rm g\,cm^{-3}}}\left(\frac{R_{\rm disk}}{R_\odot}\right)^3\left[(1+f)^3-1\right].$$

Here the factor 1.416 g cm$^{-3}$ is the mean mass density of the Sun, $3M_\odot/4\pi R_\odot^3$; we have also used the small-angle approximation $\sin\alpha \approx \alpha$. If the disk was ejected from the supernova progenitor with a velocity $v$, the mass-ejection episode that produced it is characterized by a duration $\Delta t = fR_{\rm disk}/v$ and a mass-loss rate $\dot{M} = M_{\rm disk}/\Delta t = vM_{\rm disk}/fR_{\rm disk}$.

Assuming that the disappearance of the coronal lines seen in the early-time spectra was due to the ejecta overrunning the disk (§ 3.1), we can estimate the interior disk radius $R_{\rm disk}$ using the broad-line widths. If, as we propose, these broad lines are formed in the region behind the reverse shock (§ 3.3), their velocities should be smaller than but comparable to that of the contact discontinuity between the CSM and the ejecta. At early times, the broad components of Hα, He I λ7065, and He I λ5876 all have half-widths at zero intensity of ~ 12,000 km s$^{-1}$, which yields a lower limit to $R_{\rm disk}$ of 5.7 × 10$^{15}$ cm (8.2 × 10$^4$ $R_\odot$) at day 55 when the narrow lines begin to disappear. (In comparison, the circumstellar ring around SN 1987A lies ~ 6 × 10$^{17}$ cm from the central star [Lundqvist & Fransson 1996].) If the disk was ejected from the progenitor with the velocity of the narrow lines ($v \approx 200$ km s$^{-1}$; § 3.1), this radius implies the mass-ejection episode ended >~ 9 years before the supernova explosion. We can find the fractional radius $f$ by assuming that the disappearance of all narrow lines by day 231 represents the ejecta reaching the outer radius of the disk: $f = R_{231} - R_{55}/R_{55} = 3.2$. Taking 5° = 0.087 radians as a representative value for the disk half-angle α and ρ = 1.7 × 10$^{-17}$ g cm$^{-3}$ (ρ = $n_e m_H$ if the disk is composed mainly of hydrogen; we take $n_e$ = 10$^7$ cm$^{-3}$ from § 3.3), we find $M_{\rm disk} = 0.04 M_\odot$, $\Delta t = 29$ yr, and $\dot{M} = 1.5\times10^{-3}\,M_\odot$. These values for $M_{\rm disk}$ and $\dot{M}$ are lower limits because the ejecta velocity is likely larger than that implied by the widths of the broad lines. Given that typical mass-loss rates produced by steady line-driven luminous blue variable (LBV) winds are around 10$^{-4}$ $M_\odot$ yr$^{-1}$ (Smith & Owocki 2006), these values suggest the progenitor of SN 1997eg experienced an episode of increased mass loss shortly before exploding (but note the caveats below regarding the assumptions of this simple disk model).

Alternatively, supposing that the coronal lines seen in the early-time spectra disappeared due to recombination (§ 3.1), we can infer that the ejecta had not yet completely overrun the dense line-emitting region by the time the lines disappeared (though the existence of the three distinct line components strongly suggests some circumstellar interaction, perhaps with material in the larger wind, in all our observations). Hence, the disk has a larger radius than in the previous scenario: $R_{\rm disk} > 2.4 \times 10^{16}$ cm ($3.4 \times 10^5\ R_\odot$, the radius reached by the ejecta after 231 d at 12,000 km $s^{-1}$). This implies an earlier mass ejection episode ending >~ 40 years before the explosion.

In this case we cannot calculate *f* directly, but we note that the disappearance by day 453 (Gruendl et al. 2002) of the narrow Hα P Cygni profile observed by Salamanca et al. (2002) suggests that the CSM interaction had ceased by this time. We thus take representative values for *f* between 0.01 and 2 to investigate the range of results. Table 5 shows calculated values of $M_{\rm disk}$, $\Delta t$, and $\dot{M}$ for various values of *f*, using $R_{\rm disk} = 3.4 \times 10^5\ R_\odot$ and the same values for $v$, $\rho$, and $\alpha$ as in the first scenario. Again, $M_{\rm disk}$ and $\dot{M}$ represent lower limits. The implied values of $\dot{M}$ ($10^{-3}$–$10^{-2}\ M_\odot$ $\rm yr^{-1}$) are significantly higher than the typical steady LBV mass-loss rates of $10^{-4}\ M_\odot$ $\rm yr^{-1}$, suggesting that the progenitor of SN 1997eg may have undergone a sizable eruption event at some time > 40 years before the SN explosion. The values we find for $M_{\rm disk}$ range from $10^{-3}$–$1\ M_\odot$. If the "disk" is the dense equatorial region of a larger circumstellar matter configuration, as we propose in § 5.1 above, the total mass lost in the eruption could approach 10 $M_\odot$, which Smith & Owocki (2006) give as a typical mass ejected by an LBV in such an eruption event. This exercise, while speculative, raises the possibility that the progenitor of SN 1997eg may have been an LBV, which would support the LBV–SN IIn link proposed by Gal-Yam et al. (2006) and Smith (2007). (However, see Prieto et al. 2008 for a possible counterexample.) Evidence for similar significant mass-ejection episodes a few years before explosion has been found in several previous supernovae, including SN 1994W (Chugai et al. 2004) and SN 1995G (Chugai & Danziger 2003), both of which were SNe IIn.

We note that the calculations presented here are quite sensitive to the geometrical model one assumes for the disk. Because the quantity α can be almost arbitrarily small, it dominates the scaling of $M_{\rm disk}$ and $\dot{M}$ (by contrast, $R_{\rm disk}$, ρ, and $v$ are constrained by observations and are unlikely to vary significantly from the values we have assumed here). For a very thin disk ($\alpha$ = 0.1° or 0.0017 radians) and small fractional shell thicknesses, we find minimum $\dot{M}$ values closer to the typical line-driven wind level of ~ $10^{-4}\ M_\odot$ $\rm yr^{-1}$ in both the ejecta-overrun case and the recombination case. Salamanca et al. (2002) calculate $\dot{M} = 8.3 \times 10^{-3}\ M_\odot\ \rm yr^{-1}$ from the luminosities of the Balmer emission lines, but also estimate a large circumstellar mass ($M_{\rm CSM}$ = 10 $M_\odot$). Tighter constraints on the geometrical characteristics of the disk could come from more detailed modeling of the spectropolarimetric features presented here or from a study of the light curve of SN 1997eg. Nonetheless, it is clear from the early-time behavior of SN 1997eg that the ejection of circumstellar material must have ended only recently in its history.

We can also infer that no significant deceleration of the ejecta likely occurred as a result of their expansion into the dense CSM disk (Fig. 12). The ejecta should decelerate after sweeping up a comparable mass of circumstellar material (see, e.g., McKee 1974). Our estimates of the total mass in the disk remain well below the typical ejecta mass of $\sim 10\ M_{\odot}$, even for larger values of $f$ and $\alpha$. Thus, while some slowing of the ejecta must occur to produce the double-shock structure, we do not expect that the disk caused any significant deceleration. Smith et al. (2008) reached a similar conclusion for SN 2006tf by assuming the velocity of the intermediate-width lines represented the speed of the forward shock and noting that this speed did not change over the first 100 days of the SN's evolution.

### 5.3 Origin of the Dual-Axis Structure

Dual- and multi-axis structures similar to those we propose for SN 1997eg are difficult to detect, requiring several epochs of spectropolarimetry, but a few other examples have occurred in recent years. As discussed above in § 4.2.1, Kasen et al. (2003) and Wang et al. (2003) found that the high-velocity Ca II feature in the Type Ia SN 2001el arose from a distribution of material that was kinematically and geometrically distinct from the ejecta. Wang et al. (2002) noted a 15° difference between the axes of symmetry of the ejecta and the circumstellar ring in SN 1987A, while Maund et al. (2007b) found a dramatic change in the polarization position angle between 13 and 31 days post-explosion in the Type IIb SN 2001ig. Kawabata et al. (2002) identified not two but three distinct axes within the ejecta of the Type Ic SN 2002ap.

A possibly related geometry, in which a spherically symmetric outer envelope gives way to an asymmetric core, was proposed for the Type IIb SN 1993J (Tran et al. 1997) and the Type II-P SN 2004dj (Leonard et al. 2006). Besides SN 1997eg, the only other SN IIn that has been studied in detail with spectropolarimetry is SN 1998S, which despite being a member of the same subclass displayed spectral and polarimetric behavior very different from that of SN 1997eg (Leonard et al. 2000; Wang et al. 2001). The H$\alpha$ line profile in SN 1998S was highly asymmetric, especially at later times, and had no identifiable intermediate-width component (Fig. 7 of Leonard et al. 2000). Although at very early times (15 d pre-maximum), the polarization of the broad and narrow lines departed from that of the continuum, as in SN 1997eg (Fig. 9 of Leonard et al. 2000), an observation at 42 days post-maximum showed no difference between the polarization of the broad H$\alpha$ line and its surrounding continuum (Fig. 3 of Wang et al. 2001). Relying on only the earliest epoch of polarization, Leonard et al. (2000) proposed a model very similar to the one we have developed here for SN 1997eg. However, the later data indicated that a single scattering region dominated the polarization signature, although the presence in the spectrum of narrow P Cygni line profiles led Wang et al. (2001) to suggest that some slowly moving circumstellar material still existed outside the ejecta. The pre-maximum and post-maximum polarization observations of SN 1998S occupy different regions of the $q$–$u$ plane and seem completely unrelated to each other, very much like the first two epochs of SN 2001ig (Maund et al. 2007b). As detailed supernova spectropolarimetry becomes more common, additional examples will undoubtedly emerge, and their implications for progenitor and explosion models must be addressed.

The simplest ways to explain a preferred axis in supernova ejecta are to invoke rotation of the progenitor or to posit the presence of a binary companion. In cases such as SN 1997eg and the objects discussed above, however, these explanations cannot easily account for the existence of two distinct elongation axes separated by much less than 90°. Put another way, SN 1997eg is reminiscent of the famous LBV η Car, if the central star underwent an explosion whose axis was offset from that of the homunculus/disk structure (Smith & Townsend 2007). Wang et al. (2002) suggested that the 15° offset in SN 1987A may be due to magnetic fields playing a greater role in the ejecta geometry than in the expulsion of the circumstellar ring. Maund et al. (2007a) combined features of two previous models in a "tilted-jet" picture to explain how the envelope and the core of the Type Ib/c SN 2005bf possessed different asymmetries. Perhaps the tendency of some massive stars to undergo sporadic mass ejections in random directions (e.g., VY CMa: Smith et al. 2001; Smith 2004) is also relevant in constructing the dual-axis structures seen in SN 1997eg and related objects. It is clear that more detailed observational and theoretical studies of such systems are needed and that the results of these studies will be relevant not only to SNe IIn but to supernovae of all types.

## 6. CONCLUSIONS

We have presented optical spectra of the Type IIn SN 1997eg at 12 epochs from day 16 to day 463 post-discovery, as well as spectropolarimetric data from days 16, 44, and 93. By analyzing its spectral and spectropolarimetric characteristics, we have developed a model for the system that includes an ellipsoidal or otherwise elongated helium-rich ejecta interacting with a pre-existing flattened or disk-like equatorial concentration of hydrogen-rich circumstellar material. This model is consistent with previous results that suggested the presence of dense CSM around this supernova (Salamanca et al. 2002). Our key finding is that the ejecta and the disk have different axes of symmetry; they are misaligned by ~12°, suggesting that the mass-loss phase that gave rise to the CSM and the supernova explosion process are governed by different geometrical constraints. In particular, neither the simple rotation of the progenitor star nor the presence of a companion can by itself explain the complex structure of this system.

The signatures of an equatorially enhanced CSM relate SN 1997eg to well-known objects such as SN 1987A and SN 1998S, although signatures of interaction at quite early times suggest that the equatorial material must be much closer to the supernova (by a factor of ~ 100) than is the ring of SN 1987A. The mass-loss rates implied by the density of the disk suggest that the progenitor may have been an LBV that underwent an episode of enhanced mass loss, perhaps even a significant eruption, ending roughly 10–100 years before the supernova explosion.

Most of our conclusions are independent of the interstellar polarization contribution to our data, but we prefer an ISP choice that implies an increasing intrinsic polarization with time through day 93 and an explosion date of $9^{+45}_{-9}$ days pre-discovery. The complexity of the geometrical structures we have been able to probe with three epochs of spectropolarimetry and the constraints these data have allowed us to place on our model demonstrate the wealth of information contained in such data and argue for increased spectropolarimetric observations of future supernovae, especially at multiple epochs.

We thank Alison Coil, Ron Eastman, Andrea Gilbert, Weidong Li, Maryam Modjaz, Ed Moran, and Adam Riess for assistance with the observations and data reduction, as well as Weidong Li, Peter Nugent, and Nathan Smith for helpful discussions and the anonymous referee for many thoughtful comments. J. L. H. acknowledges the support of an NSF Astronomy & Astrophysics Postdoctoral Fellowship under award AST-0302123. A. V. F.'s group at the University of California, Berkeley, is supported by NSF grant AST-0607485 and by the TABASGO Foundation. A.V.F. thanks the Lorentz Center in Leiden (the Netherlands) for its hospitality during the workshop "From Massive Stars to Supernova Remnants," when this paper was being finalized. Some of the data presented herein this paper were obtained at Lick Observatory, specifically with the Kast double spectrograph. We thank the staffs at the Keck and Lick Observatories for their assistance. Data were also obtained at the W. M. Keck Observatory, which is operated as a scientific partnership among the California Institute of Technology, the University of California, and the National Aeronautics and Space Administration. The Observatory was made possible by the generous financial support of the W. M. Keck Foundation. The authors wish to recognize and acknowledge the very significant cultural role and reverence that the summit of Mauna Kea has always had within the indigenous Hawaiian community; we are most fortunate to have the opportunity to conduct observations from this mountain.

## APPENDIX

### INTERSTELLAR POLARIZATION OF SN 1997eg

We remind the reader that in the case of extragalactic observations, ISP consists of contributions from both the Milky Way Galaxy and the host galaxy. The latter is especially difficult to quantify; recent studies have shown that the polarizing behavior of dust in other galaxies can be quite dissimilar to that of Milky Way dust (Leonard et al. 2002; Clayton et al. 2004), so one cannot assume that the well-known Serkowski relation between reddening and interstellar polarization magnitude in the Milky Way (Serkowski, Mathewson, & Ford 1975) holds for the host ISP contribution. It seems reasonable, however, to assume that like ISP in the Milky Way, ISP in other galaxies is constant with time and varies smoothly with wavelength with no sharp line effects (Jones 1989; Scarrott, Rolph, & Semple 1990), and we do so in the discussion that follows.

The two ISP contributions add vectorially in the Stokes parameters $q$ and $u$, so that at a given wavelength,

$$p_{\rm ISP,tot} = \sqrt{\left(q_{\rm ISP,MW} + q_{\rm ISP,host}\right)^2 + \left(u_{\rm ISP,MW} + u_{\rm ISP,host}\right)^2}\ ,$$

where each $q$ and each $u$ may be either positive or negative (in fact, it is possible for the host ISP partly to cancel out the Milky Way ISP). Note that while one must normally add polarized fluxes and not polarization percentages, considering only a single wavelength allows us to cancel the factor of $F_\lambda$ from both sides of this equation. Because it is unlikely that the wavelength of maximum ISP is the same in the host as in the Milky Way, the shape of the total ISP spectrum may be more complex than the simple single-peaked Serkowski curve, but we still expect that the change in polarization with wavelength of the total will be slow.

Because the Milky Way ISP contribution is constant with time and we have assumed the same for the ISP of the host galaxy NGC 5012, we know that the observed changes with time in the magnitude of continuum polarization of SN 1997eg (Table 3) must be produced by changes in the *intrinsic* polarization of the supernova. If we also assume that the continuum polarization varies over time only in magnitude and not in position angle (§ 4.1), then we can stipulate that the ISP point (which effectively defines the new origin for the ISP-corrected *q–u* diagram) should lie along the same line in *q–u* space as the continuum points; this ensures that no PA rotation occurs over time. The problem then becomes one of choosing a reasonable location for the ISP point along the continuum line.

As discussed in § 4.1, the observed changes in continuum polarization could correspond either to intrinsic increases or to intrinsic decreases, depending on the location of the ISP point. (The fact that the observed continuum polarization change is very close to linear with time strongly suggests that the intrinsic change is monotonic over the period of our observations.) Time-dependent polarimetric observations of core-collapse supernovae are still rare enough that it is difficult to make a compelling argument for either an increase or a decrease in SN 1997eg based on the behavior of similar objects. This is particularly true because the polarization in SN 1997eg appears to be due to CSM interaction as well as ejecta asymmetry, so that the polarization behavior of a SN II-P such as SN 1994dj is likely not a good basis for comparison. Among supernovae with CSM interaction, the intrinsic polarization of SN 1993J, a SN IIb that developed strong Hα emission at late times, rose from 0 to 1% between days 10 and 30 after explosion but declined slowly thereafter (at epochs comparable to those of our observations of SN 1997eg; Filippenko, Matheson, & Ho 1993; Tran et al. 1997). On the other hand, the broadband intrinsic polarization of the well known Type II SN 1987A (Jeffery 1991) increased from ~ 0.1% to ~ 1% in the first 200 days of its evolution. We must therefore consider both possibilities for SN 1997eg.

If the intrinsic continuum polarization of SN 1997eg increased with time over the period covered by our observations, then the wavelength-averaged continuum polarization we observed on day 16 (2.5%; Table 3) would represent a lower limit to the magnitude of the (total) ISP contribution at $\lambda \approx 5350$ Å. We note that the lines describing the polarization variation across the Hα and Hβ narrow emission features (Fig. 9; § 4.2) intersect the continuum axis in the region of *q–u* space described by this constraint, which we will call Case A. We can use these intersections to find a point that lies both on the axis of continuum variation and the axis of line variation. Such a point is a good ISP estimate because it simultaneously fulfills two commonly cited constraints on polarimetric data: that a constant position angle should characterize polarization variations both over time and across small wavelength ranges (Quirrenbach et al. 1997). We choose a representative ISP point between the Hα and Hβ intersections by mapping their wavelengths onto the continuum axis and locating the point between them that corresponds to 5350 Å, the wavelength average of our continuum regions. Our choice to represent Case A is $p_{\mathrm{max,\,A}} = 2.9\%$, $\theta_A = 111°$. Of the three cases we discuss here, this is our preferred choice because it ensures that both the time variation of the continuum polarization and the wavelength variation of the line polarization retain constant position angles of 35° (125°) and 23° (113°), respectively. (Harries et al. [1999] used a similar argument regarding emission-line polarization to fix the ISP contribution to the polarization of the Wolf-Rayet star EZ CMa.) In this case, the intrinsic polarization values of the narrow Balmer lines range from 0.4% to 2.5% along the 23° axis.

We digress briefly to note that this ISP choice can lead to an independent estimate of the explosion date of SN 1997eg. If we assume the intrinsic continuum polarization of SN 1997eg increased linearly with time along the continuum axis shown in Figure 6, then we can solve for the expected date at which it had zero polarization (i.e., the date at which its average continuum polarization corresponded to the ISP point we chose above for Case A). Fitting a linear relation to the relative position of the average wavelength values in Table 3 along the continuum axis as a function of time, we find $t_{\rm exp} = 9^{+45}_{-9}$ days pre-discovery, as shown graphically in Figure 11. The large uncertainties on this value reflect the distance between the Hα and Hβ intersection points in Figure 9; we have intentionally kept the estimate conservative because of the many assumptions required to obtain it. This value is roughly consistent with our previous estimate for the explosion date of 1–2 months pre-discovery based on a spectral comparison with SN 1988Z (§ 1).

If we assume instead that the intrinsic continuum polarization of SN 1997eg decreased with time between days 16 and 93 (Case B), then our observed day 93 polarization (1.5%; Table 3) represents an upper limit to the total ISP contribution. In Case B, a likely lower limit to the ISP is the point on the continuum line closest to the origin of the *q*–*u* diagram, and we will take this point as the representative ISP estimate for Case B; it occurs at $p_{\rm max,\,B} = 1.38\%$, $\theta_{\rm B} = 80°$. The intrinsic continuum polarization in this case has a PA of 125°, and the narrow Balmer emission lines have ($p$, θ) values ranging from (2.2%, 128°) to (1.3%, 161°). Both Case A and Case B are marked as asterisks in Figure 6.

Thus far we have considered the total ISP component including contributions from both the Milky Way and NGC 5012. However, because NGC 5012 lies far out of the Galactic plane, we expect the effect of Milky Way dust to be relatively small in this case. We can constrain its magnitude using Serkowski's empirical reddening relation, determined by observations of Galactic stars (Serkowski et al. 1975): $p_{\rm max}(\%) \le 9E(B-V)$. From the dust maps of Schlegel, Finkbeiner, & Davis (1998), we estimate $E(B\text{–}V)_{\rm MW} = 0.01$ mag at the position of NGC 5012, which leads to a maximum Milky Way ISP contribution of 0.09%. We checked this estimate by searching for catalogued polarimetric observations of stars near SN 1997eg at heights > 150 pc above the Galactic plane, which should serve as good probes of the Milky Way ISP (Tran et al. 1995). Four stars in the compilation by Heiles (2000) lie within 5° of SN 1997eg at the required height; they have an average polarization of (0.1 ± 0.03)%, though with widely varying position angles. An ISP contribution of this magnitude would indeed be quite small compared with the total ISP polarization values of ≥ 1.38% (previous paragraph), and we will therefore neglect the Milky Way contribution and assume $p_{\rm ISP,tot} \approx p_{\rm ISP,host}$. Though such an assumption when adding polarization vectors without knowing their position angles can lead to significant errors in the PA of the total, for $p_{\rm ISP,MW} / p_{\rm ISP,tot} < 0.1/1.38 = 0.07$, this error is ~ 2° (Hoffman et al. 2005), comparable to our instrumental PA uncertainty.

We note that in Case B described above, an ISP point farther along the continuum line in the +*u* direction is also plausible, as this would likewise result in a continuum polarization decrease with time over our observed epochs. We cannot rule out this choice, which could lead to ISP magnitudes larger than our hypothesized upper limit for Case B (albeit in a different region of *q*–*u* space). If the Serkowski relation holds in the host galaxy, then we can attempt to constrain $\rm ISP_{host}$ using the host galaxy reddening, which we estimate from the equivalent width (EW) of the blended Na I D absorption lines in the spectrum of SN 1997eg. Measuring EW(Na I)

≈ 1.3 Å in our best-quality spectra and using the bifurcated reddening relations derived by Turatto, Benetti, & Capellaro (2003), we find $0.2 < E(B\text{–}V)_{host} < 0.6$ mag, and thus $p_{max} \leq 1.8$–5.4% if the Milky Way polarization efficiency (9% mag$^{-1}$) applies to the dust in NGC 5012.

We will therefore also investigate a representative Case C, in which the ISP lies at $p = 2\%$ along the continuum line in the $+u$ region of $q$–$u$ space (also shown in Fig. 6); thus $p_{max,\,C} = 2\%$, $\theta_C = 57°$. We stress that we have no reason to expect that the dust polarization efficiency in NGC 5012 corresponds to that of the Milky Way. Clayton et al. (2004) found that the polarization efficiency for M31 along different sight lines varies over the range 7–15% mag$^{-1}$, while Leonard et al. (2002) found $31^{+22}_{-9}$ % mag$^{-1}$ for NGC 3184. However, if $p_{max}$ is significantly larger than our adopted estimate, the resulting intrinsic polarization will differ only in magnitude, not in character, from the results we derive below. We regard this Case C estimate only as a useful benchmark; it should not be interpreted as an assumption regarding the properties of the dust in NGC 5012. In Case C, the intrinsic continuum polarization has a PA of 125°, as in Case B; the narrow Balmer emission lines have ($p$, θ) values ranging from (3.6%, 126°) to (2.3%, 142°).

In all three of our ISP cases, then, the narrow lines show intrinsic polarization levels of over 2%. As discussed in § 4.3, these values contain contributions from the underlying continuum as well as the broad and intermediate-width components; due to the complexity of this situation, we have not attempted to isolate the narrow-line polarization in our analysis. The fact that the axis defined by the narrow Balmer line polarization points nearly at the origin of the $q$–$u$ plane (Fig. 9) may seem to suggest that the narrow lines are intrinsically *unpolarized*, and that they vary along the 123° axis only because of the underlying continuum and broad/intermediate line contributions. Certainly the assumption that narrow emission lines are unpolarized has been used to great advantage in the literature both of supernovae and of emission-line stars (e.g., Tran et al. 1997; Quirrenbach et al. 1997). However, although the Milky Way polarization is small along the line of sight to SN 1997eg, we cannot rule out a significant ISP contribution from the host galaxy (see above, this section), so the apparent intersection of the narrow-line axis with the origin is likely to be a coincidence. In addition, we note that any ISP choice along the narrow-line axis (including the $q$–$u$ origin) would result in an intrinsic continuum polarization that changed with time not only in magnitude but also in PA, a situation that is quite difficult to explain without invoking an extremely complex and finely tuned model.

We removed each of our three ISP estimates (Cases A, B, and C) from our observed polarization spectra for SN 1997eg, assuming $\lambda_{max} = 5350$ Å in each case to correspond to the average wavelength of our optical data (changing this parameter does not result in any significant differences in the resulting estimated intrinsic spectra). Because of space considerations, we do not present the resulting corrected spectra here, but we note that in each case the broad emission lines seen in the uncorrected polarized flux spectra (Figs. 5, 7) remain emission lines; in other words, none of these three ISP estimates implies that the broad lines represent intrinsic absorption features in the polarized flux spectra. The conclusion that broad emission lines are present in the polarization spectrum of SN 1997eg is a noteworthy consequence of our assumption that the ISP point lies on the line defined by the continuum polarization variation in the $q$–$u$ plane.

In Table 4 we present intrinsic continuum polarization measurements after removal of each of the three representative ISP estimates. To obtain these values, we performed simple vector subtraction of the ISP estimates from the values in Table 3, without taking into account the wavelength variation of the Serkowski curve; we expect any wavelength dependence to be negligible when compared with our instrumental error of 0.1%. We have not assigned any uncertainties to the ISP estimates, so the formal uncertainties on the values in Table 4 are the same as the internal errors listed in Table 3 (but we use the larger instrumental error as the uncertainty on these values). The fact that all PA values in Table 4 are nearly the same for each case reflects our assumed constraint that the ISP lies along the same line in $q$–$u$ space as the continuum observations.

Plotting estimated intrinsic polarization in each case as a function of days since discovery (Fig. 11), we see that $p$ is a nearly linear function of time and that linear least-squares fits to these points (weighted equally because the instrumental errors are constant) yield slopes with approximately the same absolute magnitude for each case. This is another consequence of our assumption that the ISP is constrained to the same line in $q$–$u$ space as the continuum polarization. Quadratic fits such as the one describing the polarization behavior of SN 2004dj (Leonard et al. 2006) not only yield significantly larger values of $\chi^2$ for Cases B and C than do the linear fits, but also predict initial polarization values (for estimated explosion dates 30–60 d pre-discovery) much higher than both the largest intrinsic supernova polarization yet observed (4.5%; Gorosabel et al. 2006) and the upper limit of ~ 4% for the representative edge-on oblate ellipsoidal scattering envelope models of Höflich (1991). (Recall from § 4.1 that we interpret the continuum polarization as arising from electron scattering in the ellipsoidal ejecta.) The expectation of a quadratic dependence is based on the fact that the optical depth of a scattering region is inversely proportional to the square of the region's size; if one assumes the region expands homologously, one infers $\tau \propto t^{-2}$. Because polarization is approximately proportional to optical depth for small optical depths (Höflich 1991; Wood et al. 1996), $p \propto t^{-2}$ if $\tau$ is small. However, as Höflich et al. note, the assumption that $p \propto \tau$ begins to break down for $\tau \approx 0.2$–$0.3$, so the fact that our observations do not follow the quadratic dependence is not surprising; it may simply indicate that the optical depth in the ejecta of SN 1997eg is large at early epochs. We observe $p \propto t$, which implies $p \propto \tau^{1/2}$, a dependence approximated by several electron-scattering models such as the edge-on prolate ellipsoidal models of Höflich (1991) at $\tau > 4$ and the thin circumstellar disks of Wood et al. (1996).

TABLE 1

JOURNAL OF OBSERVATIONS

| Day[a] | UT Date | Instrument[b] | Range (Å) | P.A.[c] (deg) | Airmass[d] | Flux Standard | Polarization Standard | Seeing (arcsec) | Grating[e] | Exposure (s) |
|---|---|---|---|---|---|---|---|---|---|---|
| 16 | 1997 Dec 20 | K2/LP | 4296–6834 | 82 | 1.11 | BD +284211 | BD +64106 | 1.2 | 600/5000 | 4 × 250 |
| 44 | 1998 Jan 17 | K2/LP | 4320–6860 | 280 | 1.04 | BD +262606 | HD 127769 | 1.2 | 600/5000 | 4 × 400 |
| 44 | 1998 Jan 17 | K2/L | 6730–10450 | 280 | 1.02 | HD 19445 | ... | 1.2 | 400/8500 | 500 |
| 55 | 1998 Jan 28 | L/K | 3280–5420 | 123 | 1.29 | Feige 34 | ... | 2.0 | 600/4310 | 1200 + 900 |
| 55 | 1998 Jan 28 | L/K | 5180–10200 | 123 | 1.29 | HD 19445 | ... | 2.0 | 300/7500 | 1200 + 900 |
| 92 | 1998 Mar 6 | K2/L | 3950–6456 | 90 | 1.29 | BD +262606 | ... | 1.1 | 600/5000 | 100 + 200 |
| 92 | 1998 Mar 6 | K2/L | 5200–8960 | 90 | 1.13 | BD +262606 | ... | 1.1 | 400/8500 | 2 × 100 |
| 93 | 1998 Mar 7 | K2/LP | 4314–6850 | 86 | 1.28 | BD +262606 | HD 155197 | 1.1 | 600/5000 | 4 × 300 |
| 113 | 1998 Mar 27 | K2/L | 3850–6350 | 130 | 1.02 | BD +262606 | ... | 0.8 | 600/5000 | 100 |
| 113 | 1998 Mar 27 | K2/L | 5440–9200 | 130 | 1.03 | HD 84937 | ... | 0.8 | 400/8500 | 100 + 30 |
| 196 | 1998 Jun 18 | L/K | 3300–5420 | 56 | 2.85 | Feige 34 | ... | 1.5 | 600/4310 | 600 + 300 |
| 196 | 1998 Jun 18 | L/K | 5100–10400 | 56 | 2.85 | HD 84937 | ... | 1.5 | 300/7500 | 600 + 300 |
| 225 | 1998 Jul 17 | L/K | 3350–5410 | 57 | 2.84 | BD +284211 | ... | 2.5 | 600/4310 | 1500 |
| 225 | 1998 Jul 17 | L/K | 5200–10400 | 57 | 2.84 | BD +262606 | ... | 2.5 | 300/7500 | 1500 |
| 231 | 1998 Jul 23 | L/K | 3280–5420 | 59 | 2.46 | BD +284211 | ... | 1.5 | 600/4310 | 1000 |
| 231 | 1998 Jul 23 | L/K | 4300–7050 | 59 | 2.14 | BD +262606 | ... | 1.5 | 600/5000 | 1000 |
| 231 | 1998 Jul 23 | L/K | 6050–8050 | 59 | 2.47 | BD +262606 | ... | 1.5 | 830/8460 | 1000 |
| 356 | 1998 Nov 25 | L/K | 3330–5510 | 121 | 1.37 | BD +284211 | ... | 2.2 | 600/4310 | 900 |
| 356 | 1998 Nov 25 | L/K | 5200–10200 | 121 | 1.37 | BD +174708 | ... | 2.2 | 300/7500 | 900 |
| 402 | 1999 Jan 10 | L/K | 3350–5500 | 180 | 1.04 | BD +284211 | ... | 1.5 | 600/4310 | 1500 + 1100 |
| 402 | 1999 Jan 10 | L/K | 5200–10300 | 180 | 1.04 | BD +174708 | ... | 1.5 | 300/7500 | 1500 + 1100 |
| 463 | 1999 Mar 12 | L/K | 3276–5400 | 185 | 1.29 | Feige 34 | ... | 2.4 | 600/4310 | 1800 |
| 463 | 1999 Mar 12 | L/K | 5250–10500 | 185 | 1.29 | HD 19445 | ... | 2.4 | 300/7500 | 2100 |

[a] Days since discovery, 1997 December 4 UT (HJD 2,450,787).
[b] L/K = Lick 3 m/Kast Double Spectrograph; K2/L = Keck II 10 m/Low Resolution Imaging Spectrometer (LRIS); K2/LP = Keck II 10 m/LRIS with Polarimeter.
[c] Position angle of the spectrograph slit. All observations were taken within 5° of the parallactic angle (Filippenko 1982), except those on days 16, 44, 113, and 402, which were observed at low airmass (§ 2).
[d] Airmass at midpoint of exposure(s).
[e] Grating denoted by number of lines $mm^{-1}$/blaze wavelength (Å). Typical grating resolutions: ~ 12 Å for 300/7500, ~ 8 Å for 400/8500, ~ 8 Å for 600/4310, ~ 6 Å for 600/5000 (Keck K2/L), ~ 10 Å for 600/5000 (Keck K2/LP), and ~ 7 Å for 830/8460.

TABLE 2

EMISSION-LINE IDENTIFICATIONS AND TIME EVOLUTION

| Line | Day 16 | Day 44 | Day 55 | Day 92[a] | Day 113 | Day 196 | Day 225 | Day 231 | Day 356 | Day 402 | Day 463 | Comments |
|---|---|---|---|---|---|---|---|---|---|---|---|---|
| O III 3341 | ... | ... | n | ... | ... | ... | ... | ... | ... | ... | ... | Bowen line |
| Ca II 3347 | ... | ... | b | ... | ... | ... | ... | ... | ... | ... | ... | blended with narrow Fe I? 3355 |
| [Ne V] 3426 | ... | ... | n | ... | ... | ... | ... | ... | ... | ... | ... | ... |
| O III 3444 | ... | ... | n | ... | ... | ... | ... | ... | ... | ... | ... | Bowen line |
| Broad feature 3800–4000 | ... | ... | flat | ... | symm | symm | symm | symm | ... | ... | ... | Ca II / He I blend |
| He I 3820 | ... | ... | n | ... | ... | ... | ... | ... | ... | ... | ... | ... |
| Hη 3835 | ... | ... | ... | ... | PC | PC | ... | ... | n? | ... | n? | ... |
| [Ne III] 3869 | ... | ... | n | ... | n | ... | ... | ... | ... | ... | ... | ... |
| He I / Hζ 3889 | ... | ... | PC | ... | PC? | ... | ... | ... | ... | ... | ... | ... |
| Ca II K 3934 | ... | ... | a | a | a | a | n? | ... | ... | ... | ... | ... |
| He I 3965 / [Ne III] 3967 | ... | ... | n | ... | ... | ... | ... | ... | ... | ... | ... | ... |
| Ca II H 3969 / Hε 3970 | ... | ... | ... | ... | a | ... | ... | ... | ... | ... | ... | ... |
| S II 4069 | ... | ... | n | n | ... | ... | ... | ... | ... | ... | ... | ... |
| Hδ 4102 | ... | ... | b+i+n | b+i+n | b+dbl | b+i | b+i | b+i | ... | ... | ... | int component blueshifted at early epochs; narrow component moves from blue to red |
| Ca I 4227 | ... | ... | | b | ... | b | b | ... | ... | ... | ... | stronger in later two spectra; ID uncertain |
| Hγ 4340 | b+i+n | b+i+n | b+i+n | b+dbl? | b+dbl? | b+flat | b+flat | b+flat | b+i | b+i+n | b+i+n? | broad line blueshifts with time; int component blueshifted at early epochs |
| [O III] 4363 | n | n | n | n | n | n | n | n | n? | ... | ... | strength decreases with time |
| He I 4472 | n | n | ... | ... | ... | ... | ... | ... | ... | ... | ... | ... |
| Broad feature 4430–4650 | ... | b | b | b | b | b | b | b | b | b? | ... | strongest at middle epochs, slightly redshifts with time; Ba II 4544? |
| Fe II / [Fe III] 4658 | n | n | n | n | n | n? | ... | ... | ... | ... | ... | weakens with time |

[a] The spectrum from day 93 is sufficiently similar to that from day 92 that we do not include it in this table.

| Line | Day 16 | Day 44 | Day 55 | Day 92[a] | Day 113 | Day 196 | Day 225 | Day 231 | Day 356 | Day 402 | Day 463 | Comments |
|---|---|---|---|---|---|---|---|---|---|---|---|---|
| He II 4686 | n | n | n | n | n | ... | ... | ... | ... | ... | ... | weakens with time |
| Hβ 4861 | b+i+n | b+i+n | b+i+n | b+i+n | b+i+n | b+dbl | b+dbl | b+dbl | b+i+n | b+i+n | b+i+n | narrow component weakens; int component blueshifted at early epochs; broad component narrows |
| [Fe IV] 4906 | n | n | ... | ... | ... | ... | ... | ... | ... | ... | ... | ... |
| [O III] 4959 | n | n | n | n | n | ... | ... | ... | n? | n? | ... | ... |
| [O III] 5007 | n | n | n | n | n | ... | ... | ... | ... | n | n | ... |
| He I 5017 | b+n | b+n? | b | b | b | b | b | b | flat? | b+n? | n | broad component strengthens in mid-epochs |
| Si II 5041 | n | n | n | n | ... | ... | ... | ... | ... | ... | ... | ... |
| Broad feature 5100–5400 | b | b | b | b | b | b | b | b | b | b | ... | strengthens in mid-epochs; likely a blend of Fe II and other iron lines |
| [Fe VII] 5159 / [Fe II] 5160 | n | n | n | n? | ... | ... | ... | ... | ... | ... | ... | ... |
| N II / [Fe VI] 5176 | n | n | n | n | ... | ... | ... | ... | ... | ... | ... | ... |
| Blend 5232 | n | n | n | n | n | ... | ... | ... | ... | ... | ... | ... |
| Blend 5270 | n | n | n | n | n | ... | n | ... | ... | ... | ... | Fe II 5269 / [Fe III] 5270 / Fe II 5273? |
| [Fe XIV] 5303 / [Ca V] 5309 | n | n | n? | n | n | ... | ... | ... | ... | ... | ... | ... |
| He II 5411 | n | n | n? | n? | ... | ... | ... | ... | ... | ... | ... | ... |
| [Fe VI] / Fe I 5429 | n | n | ... | n | ... | ... | ... | ... | ... | ... | ... | ... |
| [Ar X] 5535 | n | n | ... | n | n | ... | ... | ... | ... | ... | ... | ... |
| [Fe VII] 5720 | n | n | ... | ... | ... | ... | ... | ... | ... | ... | ... | ... |
| [N II] 5755 | n | n | n | n | n | ... | ... | n? | ... | ... | ... | ... |
| He I 5876 | b+n | b+n | b | b | b | b | b | b | b+n? | b+n? | b+a | broad line triangular, blueshifts with time; symm in middle epochs |
| [Fe VII] 6086 | n | n | ... | ... | b? | ... | ... | ... | ... | ... | ... | ... |
| Fe II 6258 band | b | b | ... | ... | ... | ... | ... | ... | b? | ... | ... | ... |
| [Fe X] 6374 | n | n | n | n | n | n | ... | n | ... | ... | ... | ... |
| Hα 6563 | b+i+n | b+i+n | b+i+n | b+i+n | b+i+n | b+i | b+i | b+dbl | b+i | b+i+n | b+i+n | int component blueshifted at early epochs; multiple components at late times? |
| He I 6678 | b+n | b | b | b? | ... | ... | ... | ... | ... | ... | ... | broad line triangular, becomes dbl-peaked; blended with He I 7281? |
| [S II] 6716 | ... | ... | n | ... | n? | n? | n | n | n? | n | n | strengthens with time |

| Line | Day 16 | Day 44 | Day 55 | Day 92[a] | Day 113 | Day 196 | Day 225 | Day 231 | Day 356 | Day 402 | Day 463 | Comments |
|---|---|---|---|---|---|---|---|---|---|---|---|---|
| [S II] 6732 | ... | ... | n? | ... | n? | ... | n | n | n? | n | n | strengthens with time |
| He I 7065 | ... | b+n | b+n | b+n | b+n | b+n | b+n | b | b? | b+n? | b | ... |
| He I 7281 | ... | n | b? | b? | b? | b | b | b | ... | ... | b?+n | ... |
| [O II] 7320 | ... | n | n | n | n | ... | ... | n? | ... | n? | ... | ... |
| [O II] 7330 | ... | n | n | n | ... | ... | ... | ... | ... | n? | ... | ... |
| [Fe XI] 7892 | ... | n | n | n | n | n? | ... | ... | ... | ... | ... | ... |
| O I 8446 | ... | n | n | n | n | n | n? | ... | n | n | n? | blended with [Ca II] triplet |
| [Ca II] 8498 | ... | b | b | b | b | b | b | ... | a? | ... | ... | blended with other [Ca II] lines |
| [Ca II] 8542 | ... | ... | ... | ... | ... | n? | ... | ... | n? | ... | ... | blended with other [Ca II] lines |
| [Ca II] 8662 | ... | b | b | b | b | b | b | ... | ... | ... | ... | blended with other [Ca II] lines |
| Pa11 8863 | ... | ... | ... | ... | ... | ... | ... | ... | ... | n? | n? | emission seems redshifted |
| Pa10 9014 | ... | b+n | b? | ... | b | a? | b? | ... | n? | ... | ... | ... |
| Pa9 9229 | ... | b+n | b+n? | ... | ... | a | b+n? | ... | ... | ... | ... | ... |

NOTE.— Lines are described as follows: "a" = absorption; "b" = broad; "dbl" = double-peaked intermediate-width component; "flat" = flat-topped intermediate-width component; "i" = intermediate-width; "n" = narrow; "PC" = P Cyg profile; "symm" = symmetric about line center. A question mark indicates an uncertain detection due to noise in the spectrum. For spectra that do not cover the entire wavelength range shown here, horizontal rules indicate their deredshifted wavelength limits.

TABLE 3

OBSERVED CONTINUUM POLARIZATION AND POSITION ANGLE OF SN 1997EG

| | $p$, $\sigma_p$ (%)[a] | | | $\theta$, $\sigma_\theta$ (°)[a] | | |
|---|---|---|---|---|---|---|
| | Day 16 | Day 44 | Day 93 | Day 16 | Day 44 | Day 93 |
| Blue (4500–4600 Å) | 2.63 ± 0.09 | 2.09 ± 0.06 | 1.49 ± 0.06 | 110.5 ± 0.9 | 103.6 ± 0.8 | 96.1 ± 1.1 |
| Green (5400–5500 Å) | 2.37 ± 0.08 | 2.22 ± 0.05 | 1.59 ± 0.06 | 107.2 ± 1.0 | 102.7 ± 0.7 | 92.7 ± 1.1 |
| Red (6100–6200 Å) | 2.55 ± 0.07 | 2.11 ± 0.06 | 1.52 ± 0.06 | 110.9 ± 0.7 | 105.1 ± 0.9 | 97.3 ± 1.1 |
| Wavelength average (5350 Å) | 2.51 ± 0.04 | 2.14 ± 0.03 | 1.53 ± 0.03 | 109.6 ± 0.5 | 103.8 ± 0.5 | 95.3 ± 0.6 |

NOTE.—Epochs refer to number of days post-discovery. Averages were carried out in Stokes $q$ and $u$.

[a] Uncertainties given are internal measurement errors, propagated to reflect binning over wavelength. We estimate uncertainties due to instrumental polarization to be $\sigma_{p,\,\mathrm{inst}} = 0.1\%$ and $\sigma_{\theta,\,\mathrm{inst}} = 2°$ for all observations.

TABLE 4

CONTINUUM POLARIZATION AND POSITION ANGLE OF SN 1997EG CORRECTED FOR ISP

| Case A[a] | *p* (%)[b] | | | θ (°)[b] | | |
|---|---|---|---|---|---|---|
| | Day 16 | Day 44 | Day 93 | Day 16 | Day 44 | Day 93 |
| Blue (4500–4600 Å) | 0.3 | 1.0 | 1.8 | 28 | 37 | 34 |
| Green (5200–5500 Å) | 0.6 | 1.0 | 1.9 | 37 | 41 | 36 |
| Red (6100–6200 Å) | 0.3 | 0.9 | 1.7 | 23 | 35 | 33 |
| Wavelength average (5350 Å) | 0.4 | 1.0 | 1.8 | 31 | 38 | 35 |
| Case B[a] | | | | | | |
| Blue (4500–4600 Å) | 2.3 | 1.5 | 0.8 | 126 | 124 | 129 |
| Green (5200–5500 Å) | 1.9 | 1.6 | 0.7 | 125 | 122 | 123 |
| Red (6100–6200 Å) | 2.3 | 1.6 | 0.9 | 127 | 126 | 129 |
| Wavelength average (5350 Å) | 2.2 | 1.6 | 0.8 | 126 | 124 | 127 |
| Case C[a] | | | | | | |
| Blue (4500–4600 Å) | 3.7 | 3.0 | 2.2 | 126 | 125 | 127 |
| Green (5200–5500 Å) | 3.4 | 3.0 | 2.1 | 125 | 123 | 124 |
| Red (6100–6200 Å) | 3.7 | 3.1 | 2.3 | 126 | 125 | 127 |
| Wavelength average (5350 Å) | 3.6 | 3.0 | 2.2 | 126 | 124 | 126 |

NOTE.—The values presented here were obtained by subtracting each of three representative ISP estimates (§ 4.3) from the values in Table 3. We have neglected the wavelength dependence of the Serkowski curve, which we expect to be small over the spectral region represented here. Epochs refer to number of days post-discovery.

[a] Our three ISP estimates are as follows. Case A: $p_{max}$ = 2.9%; θ = 111°; Case B: $p_{max}$ = 1.4%; θ = 80°; Case C: $p_{max}$ = 2%; θ = 57°.

[b] We have not assigned any formal errors to the ISP estimates; instead we take the uncertainties on the values tabulated here to be those due to instrumental polarization, which we estimate to be $\sigma_{p,\,inst}$ = 0.1% and $\sigma_{\theta,\,inst}$ = 2° for all observations.

TABLE 5

ESTIMATED MASS-LOSS PROPERTIES OF THE SN 1997EG PROGENITOR

| $f$[a] | $M_{disk}$[b] ($M_\odot$) | $\Delta t$ (yr)[c] | $\dot{M}$[d] ($M_\odot$ yr$^{-1}$) |
|---:|---:|---:|---:|
| 0.01 | 0.001 | 0.38 | $3.42 \times 10^{-3}$ |
| 0.05 | 0.007 | 1.90 | $3.56 \times 10^{-3}$ |
| 0.1 | 0.014 | 3.80 | $3.74 \times 10^{-3}$ |
| 0.5 | 0.010 | 19.02 | $5.36 \times 10^{-3}$ |
| 1 | 0.301 | 38.05 | $7.90 \times 10^{-3}$ |
| 2 | 1.12 | 76.09 | $1.47 \times 10^{-2}$ |

NOTE.—These calculations assume the ejecta had not overrun the narrow-line emitting region by day 231. For purposes of illustration, we have taken the radius of the disk-like circumstellar shell to be $3.4 \times 10^5\ R_\odot$, its expansion velocity to be 200 km s$^{-1}$, its density to be $1.7 \times 10^{-17}$ g cm$^{-3}$, and its opening angle to be 5° (see § 5.2 for further details). Increasing any of these values causes $\dot{M}$ to increase.

[a] We define the fractional thickness of the disk-like circumstellar shell to be $f = \Delta R/R_{min}$; see § 5.2.

[b] $M_{disk}$ values represent lower limits to the total mass of the circumstellar shell.

[c] We define the duration of the mass-loss episode that formed the circumstellar shell to be $\Delta t = fR_{min}/v$; see § 5.2.

[d] $\dot{M}$ values represent lower limits to the mass loss rate of the episode that formed the circumstellar shell.

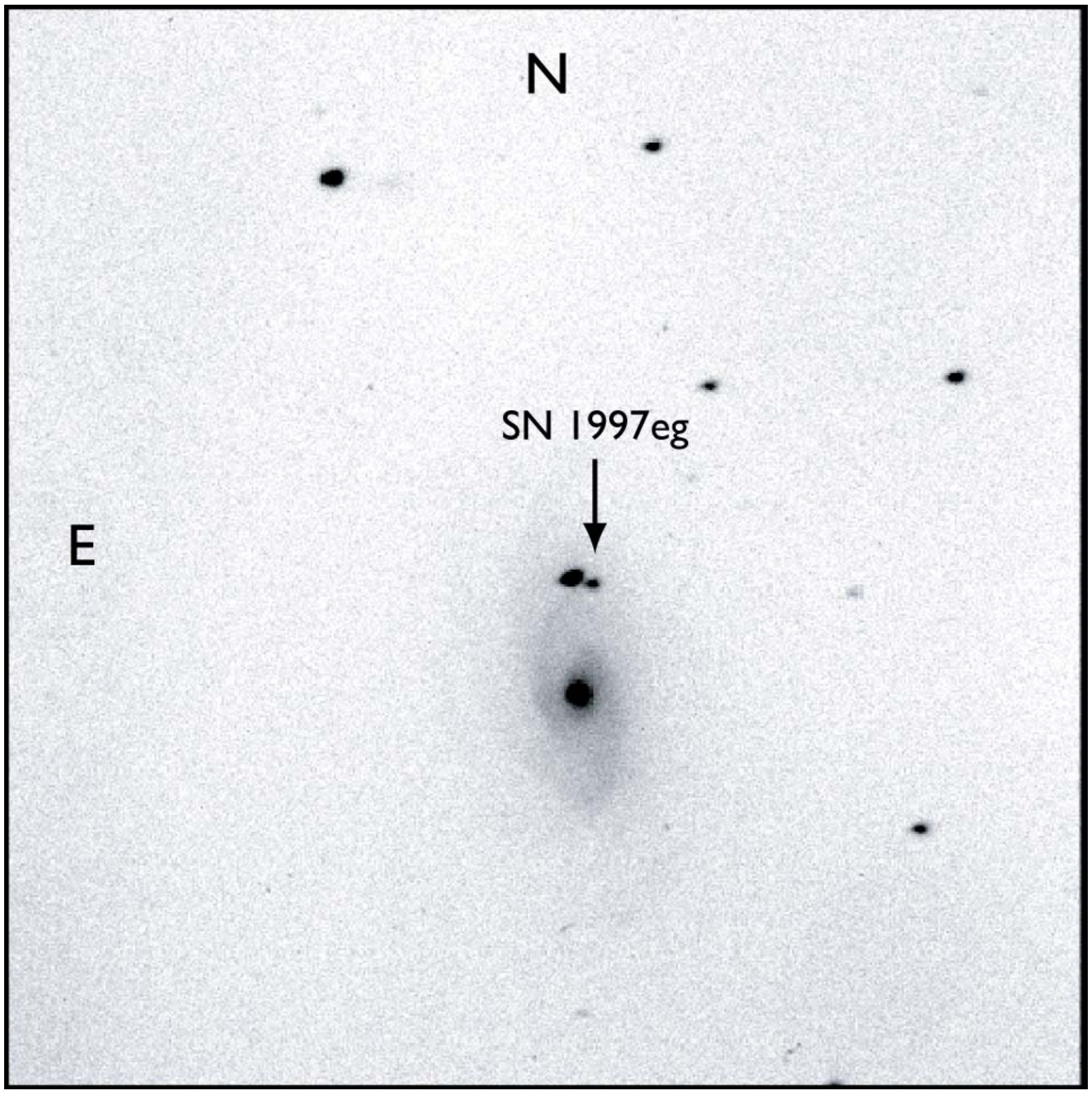

FIG. 1—*I*-band image of NGC 5012 and SN 1997eg taken with KAIT (Filippenko et al. 2001) on 1998 January 22. The field of view is 5.4′ × 5.4′. The supernova was located about 5″ west of a foreground star of magnitude 13.9 (Nakano & Aoki 1997).

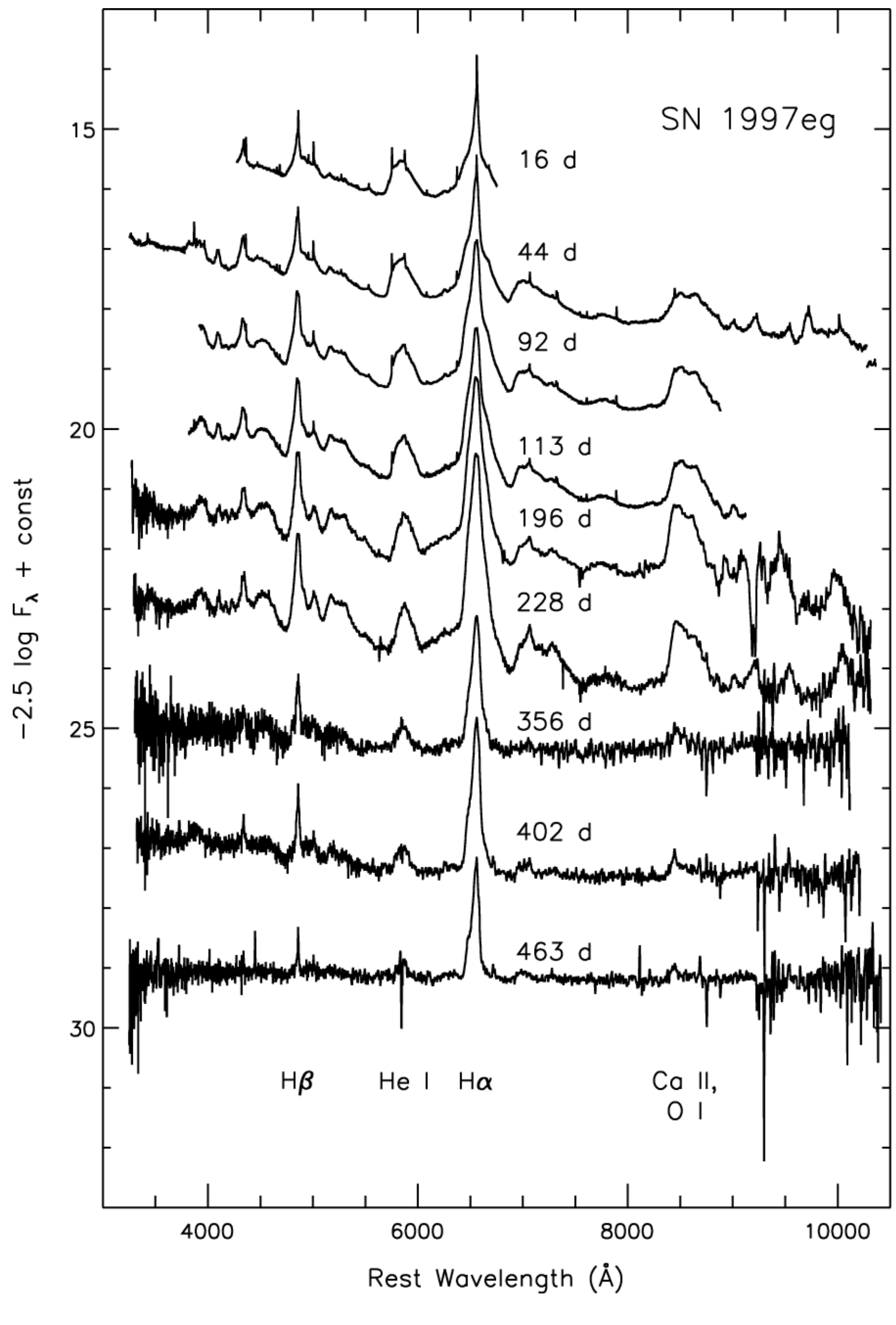

FIG. 2—Montage of spectra of SN 1997eg with prominent features identified. Spectra have been corrected for the host-galaxy recession velocity, $v = 2485$ km s$^{-1}$, determined from the early-time narrow-line features (in agreement with Salamanca et al. 2002). Epochs (days) are given relative to the discovery date, 1997 December 4. The day 44 spectrum contains the region 3254–4284 Å from a bluer spectrum taken on day 55 (not shown). The spectrum labeled "92 d" is the average of spectra taken on days 92 and 93, and the spectrum labeled "228 d" is the average of spectra taken on days 225 and 231.

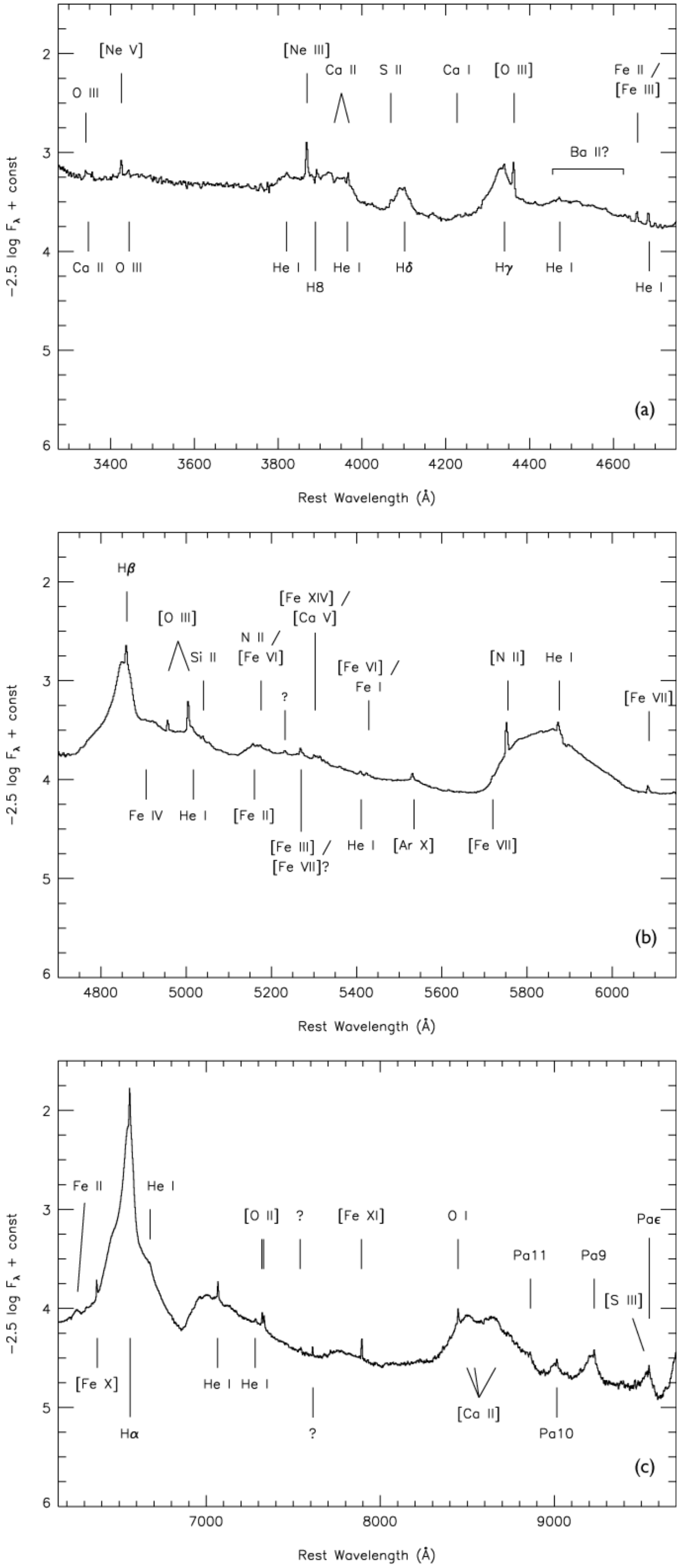

FIG. 3—(*a*) Detailed early-time spectrum (composite of spectra from day 44 and day 55 post-discovery) of SN 1997eg over the range 3300–4700 Å with prominent lines identified (see Table 2 for wavelengths and line identifications). Spectra have been corrected for the host-galaxy recession velocity as in Fig. 2. (*b*) As in (*a*), but for 4700–6100 Å. (*c*) As in (*a*), but for 6100–9700 Å.

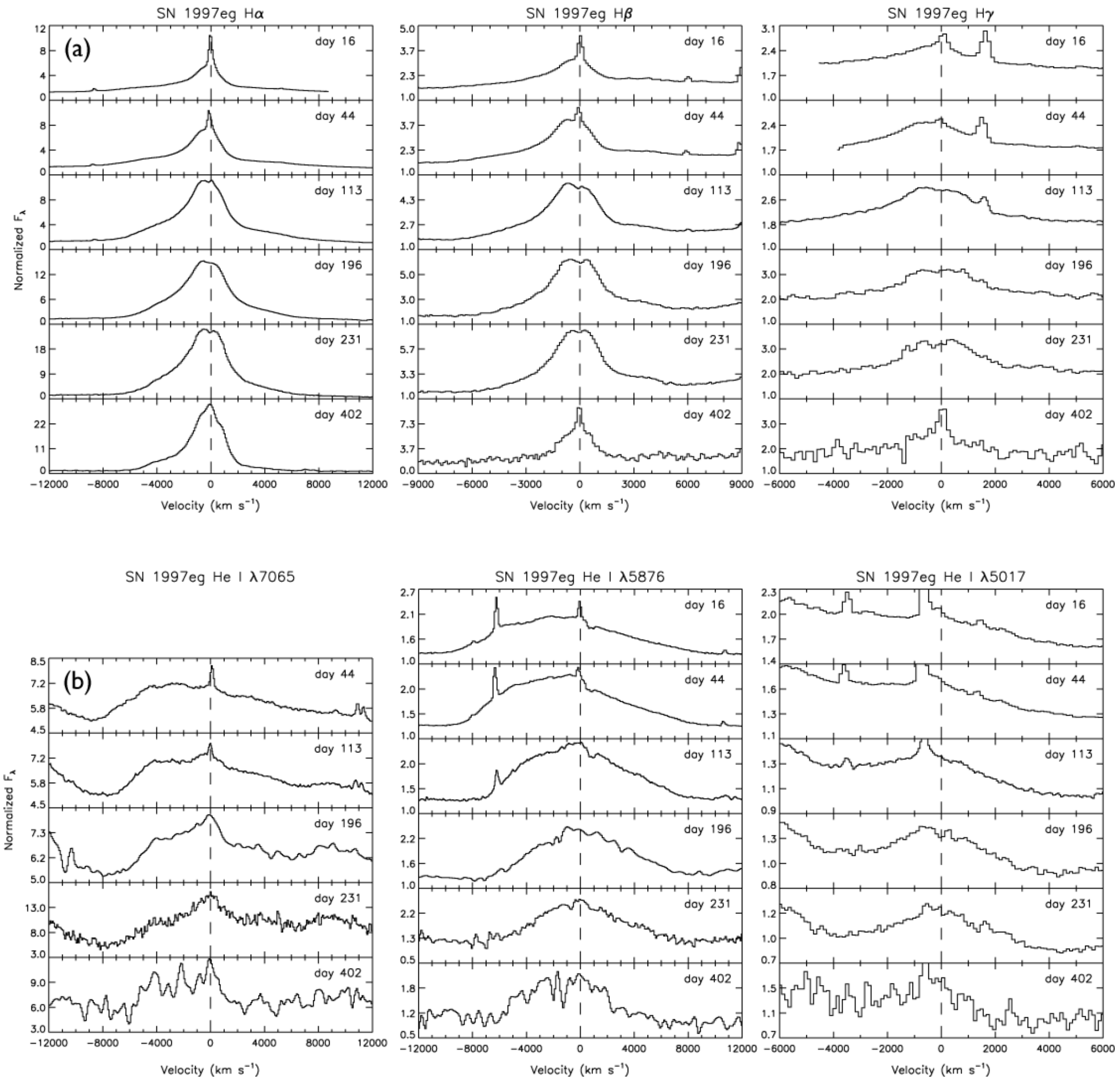

FIG. 4—(*a*) Evolution of the multi-component Hα, Hβ, and Hγ profiles of SN 1997eg with time. Spectra are normalized to day 16 at $v = -10{,}000$ km s$^{-1}$, –8000 km s$^{-1}$, and –4000 km s$^{-1}$, respectively, but are displayed on different scales in each panel for comparison of the line shapes. The narrow line on the red side of Hγ in the early-time spectra is [O III] λ4363. (*b*) As in (*a*), but for the He I λλ7065, 5876 and 5017 profiles. The He I λ5876 and λ5017 spectra are normalized to day 16 at $v = -10{,}000$ km s$^{-1}$ and –6,000 km s$^{-1}$, respectively; the He I λ7065 spectrum is normalized to day 44 at –8000 km s$^{-1}$. The narrow line on the blue side of He I λ5876 in the early-time spectra is [N II] λ5755; the strong narrow lines on the blue side of He I λ5017 are [O III] λλ4959, 5007. We have no data for He I λ7065 on day 16 because that spectrum extends only to 6834 Å (Table 1).

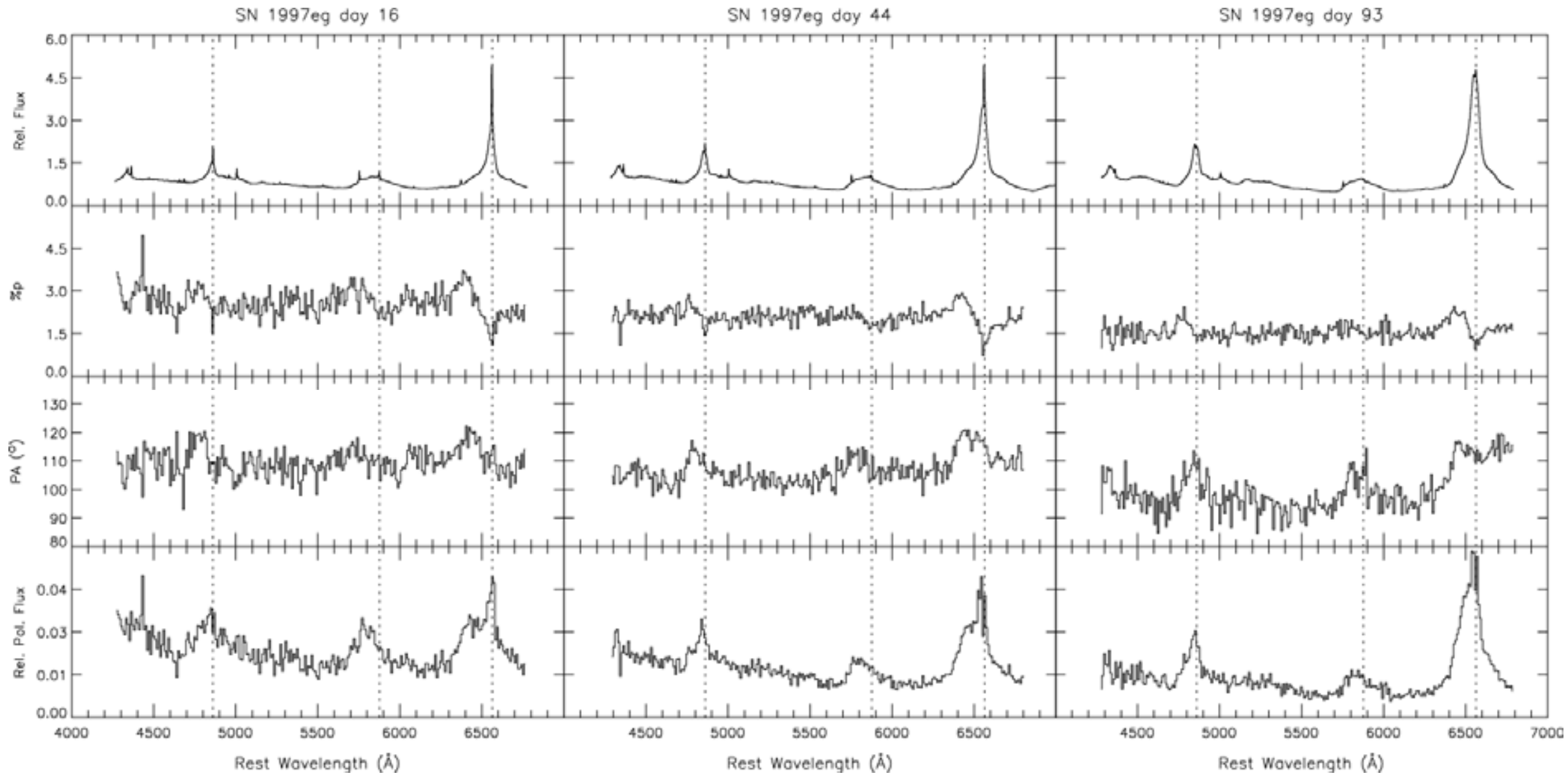

FIG. 5—Relative total-flux, percent polarization, position angle (PA), and relative polarized flux ($p \times F_\lambda$) spectra for SN 1997eg on day 16, day 44, and day 93 post-discovery. For clarity, flux spectra have been normalized to 1 at 4500 Å and the bottom three spectra have been binned to a resolution of 10 Å. All spectra have been corrected for the host-galaxy recession velocity as in Fig. 2. Vertical dotted lines mark the rest wavelengths of Hα, Hβ, and He I λ5876.

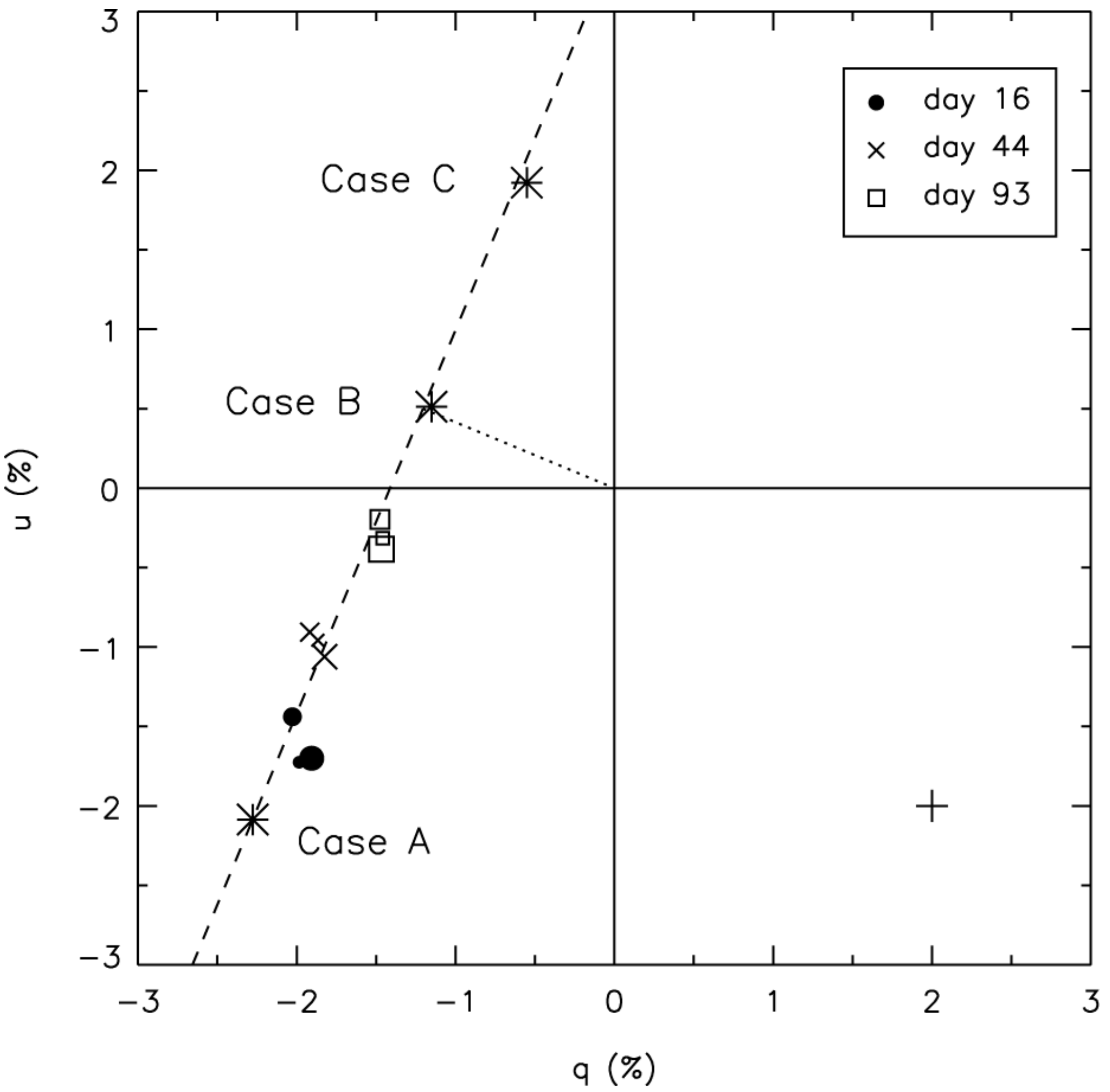

FIG. 6—Continuum polarization measurements of SN 1997eg (Table 3) plotted in the Stokes *q–u* plane; filled circles represent day 16 data, "X" symbols represent day 44 data, and open squares represent day 93 data. Symbol size is proportional to wavelength, so that the smallest symbols represent the "blue" continuum region and the largest symbols represent the "red" continuum region. The dashed line shows a least-squares fit, weighted by internal errors (not shown here), to the nine continuum points; the equation of the line is $u = 2.9q + 4.2$, corresponding to a PA of 35° or 125° (§ 4.1). The cross in the lower right quadrant represents the systematic instrumental uncertainty on each of the continuum points. Asterisks mark our three interstellar polarization estimates (§ 4.3; Appendix).

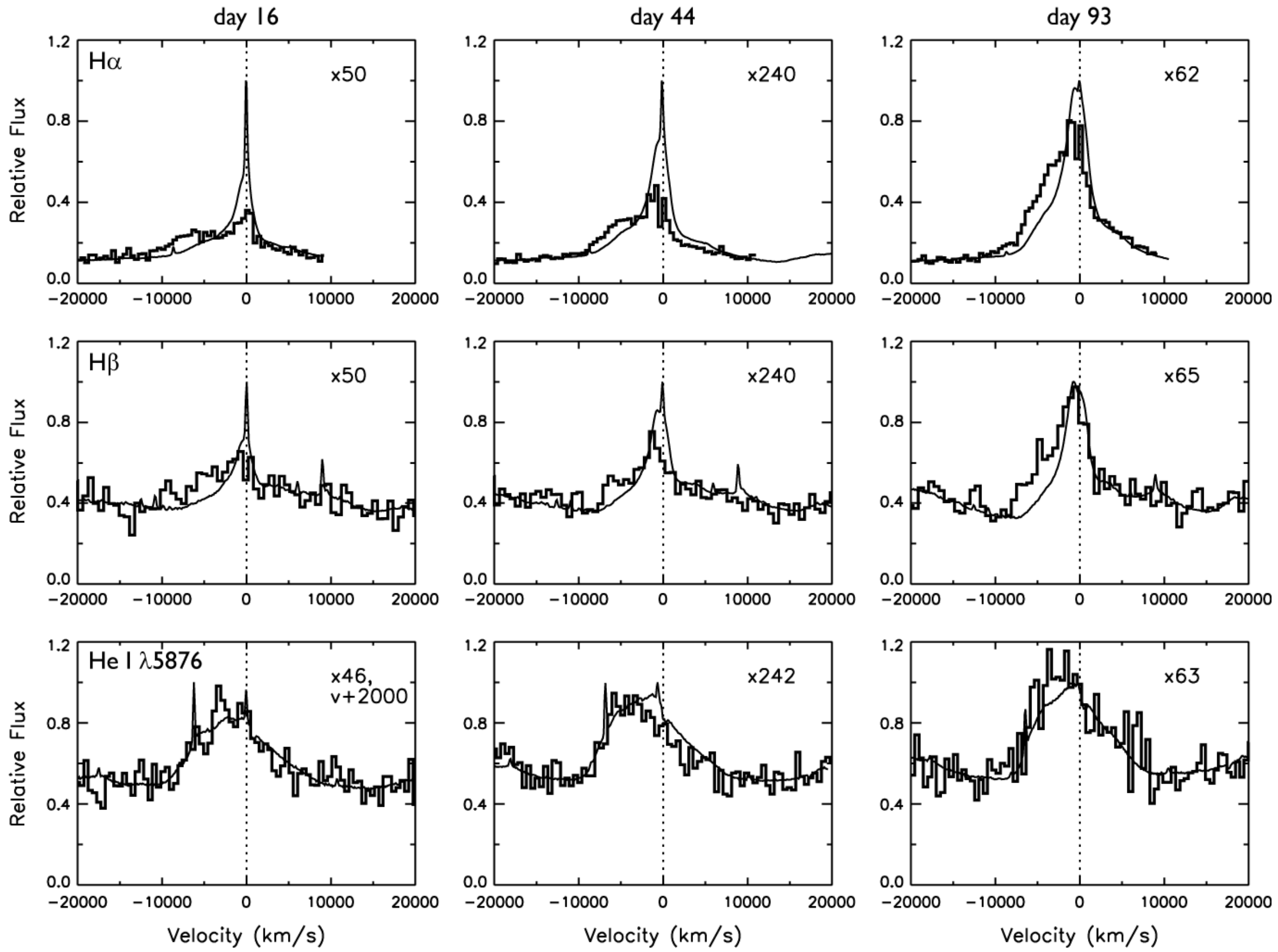

FIG. 7—Line profiles of the Hα (*top row*), Hβ (*middle row*), and He I λ5876 (*bottom row*) emission lines in total flux (*narrow smooth lines, left-hand axes*) and polarized flux (*thick binned lines, right-hand axes*) at each of our three epochs of spectropolarimetry (*left to right:* days 16, 44, and 93 post-discovery). Total-flux spectra have been normalized to their respective line peaks. Each polarized flux spectrum has been binned to a resolution of 10 Å and multiplied by the same normalizing factor as its corresponding total-flux spectrum, then by an additional factor shown in each frame to facilitate direct comparison of the line shapes. In the first epoch, the polarized flux line of He I λ5876 has also been redshifted by 2000 km $s^{-1}$. Dotted lines represent zero velocity.

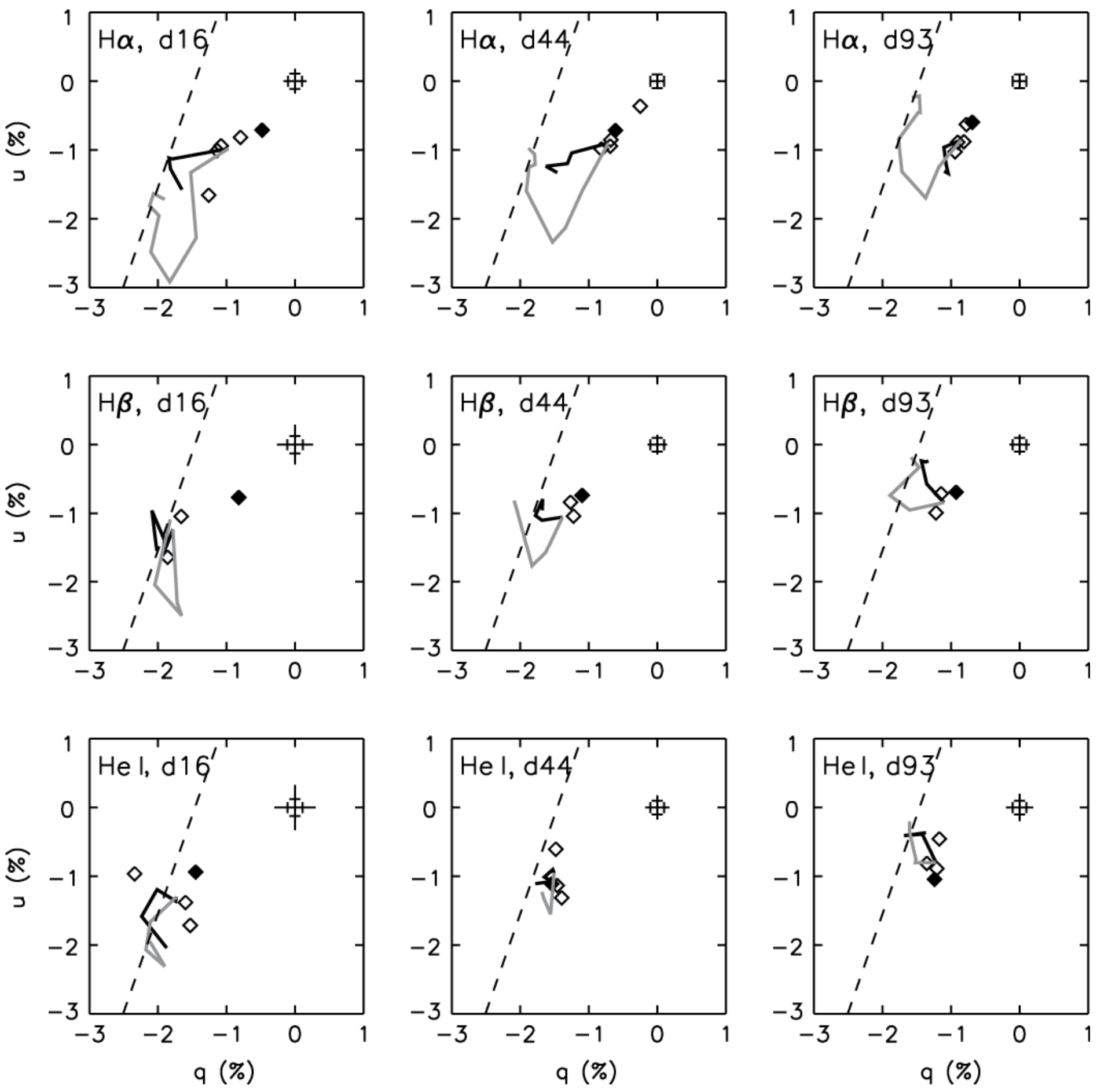

FIG. 8—Polarization profiles of the Hα (*top row*), Hβ (*middle row*), and He I λ5876 (*bottom row*) emission lines in the Stokes *q*–*u* plane for each of our three epochs of spectropolarimetry (*left to right*: days 16, 44, and 93 post-discovery). Data plotted as connected lines represent the broad-line components and have been binned to a resolution of 50 Å, with gray lines representing negative (blueshifted) velocities and black lines representing positive (redshifted) velocities. Data for Hα range from –18,200 to +9000 km $s^{-1}$; they are asymmetric around line center due to the LRIS-P red-wavelength cutoff. Data for Hβ cover $|v| < 12,220$ km $s^{-1}$ and data for He I λ5876 cover $|v| < 10,100$ km $s^{-1}$. Narrow-line polarization ($|v| < 550$ km $s^{-1}$) is overplotted in bins of width 5 Å (*diamonds*); the rest wavelength is shown as a filled diamond in each panel. Each straight dashed line represents the fit to the continuum polarization variation with time, as in Fig. 6. The crosshair symbol in each panel both marks the origin of the plot and shows the size of the uncertainties; the outer arms of each cross represent the average error bars on the diamond points, while the inner hash marks represent the average error bars on the points that make up the gray and black lines. In both cases, the error bars are taken to be the larger of the internal uncertainties and the instrumental uncertainty of 0.1%.

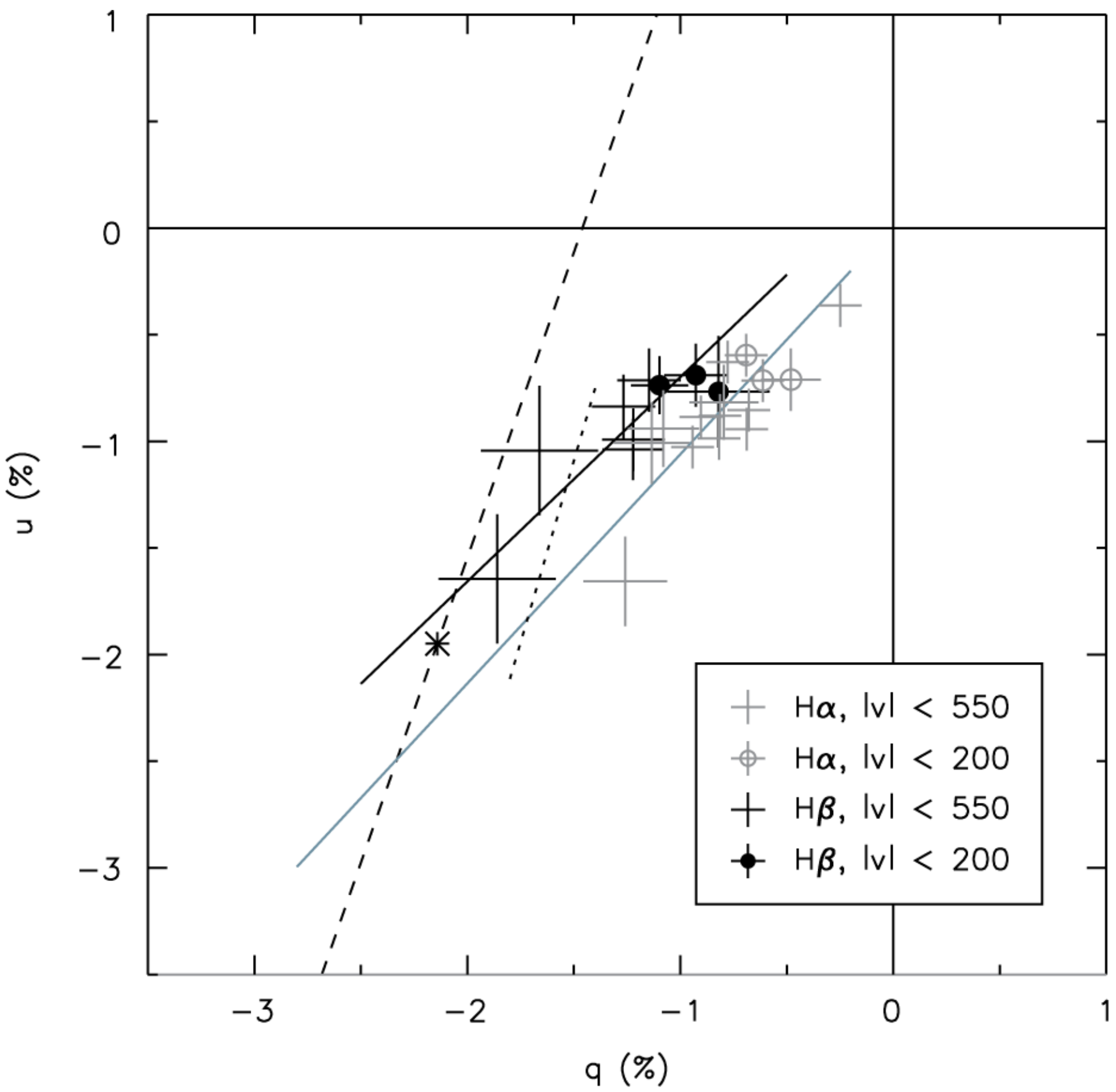

FIG. 9—Polarization of the narrow lines ($|v| < 550$ km s$^{-1}$) of Hα (*gray lines, open circles*) and Hβ (*black lines, filled circles*) in the Stokes *q*–*u* plane; all three epochs are shown together. Each cross represents the polarization and associated uncertainties in a bin 5 Å wide. Internal uncertainties are shown unless they are smaller than the instrumental uncertainty of 0.1%. Circles represent line regions with $|v| < 200$ km s$^{-1}$. Diagonal lines represent least-squares fits, weighted by internal errors, to the Hα and Hβ points; their equations are $u = 1.11q + 0.01$ and $u = 0.97q + 0.27$, respectively. The short dotted line represents a similar fit to the polarization of the narrow line of He I λ5876; its equation is $u = 10.91q + 14.47$. The dashed line is the fit to the continuum polarization variation with time (Fig. 6). The star represents the point we take for our "Case A" ISP estimate (Fig. 6; § 4.3; Appendix).

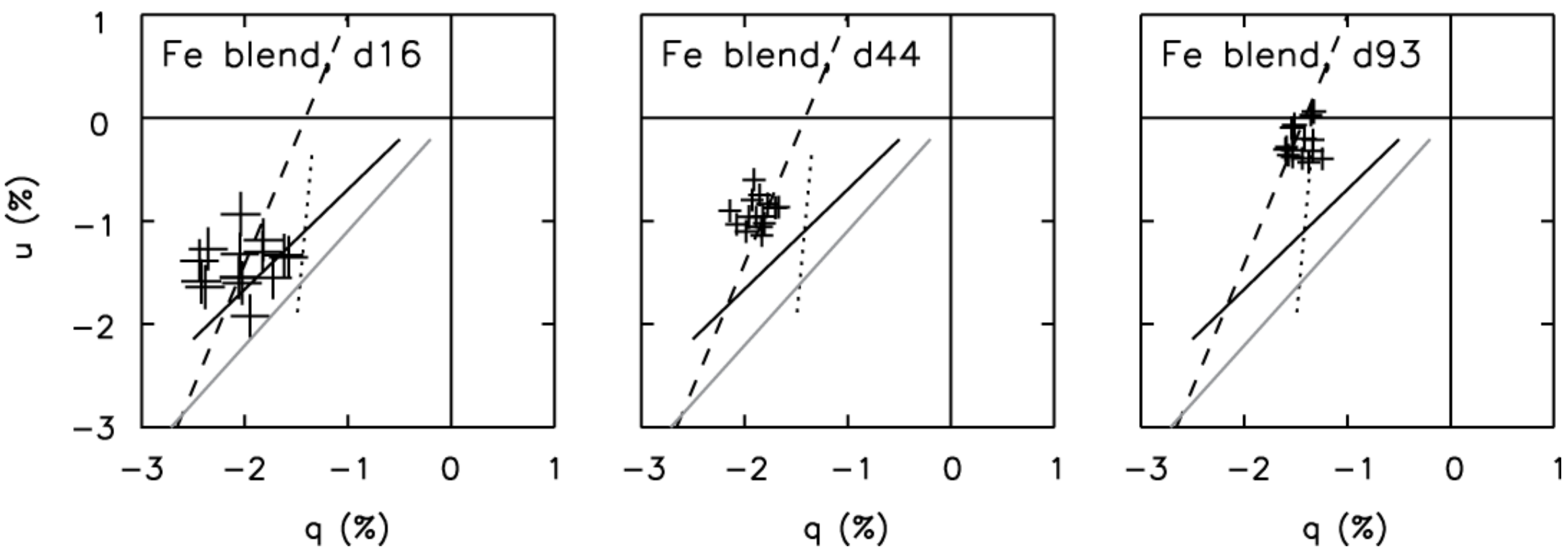

FIG. 10—Polarization of the broad iron blend at 5100–5400 Å. Each cross represents the polarization and associated uncertainties in a bin 20 Å wide. Black, gray, dotted, and dashed lines are as in Fig. 9.

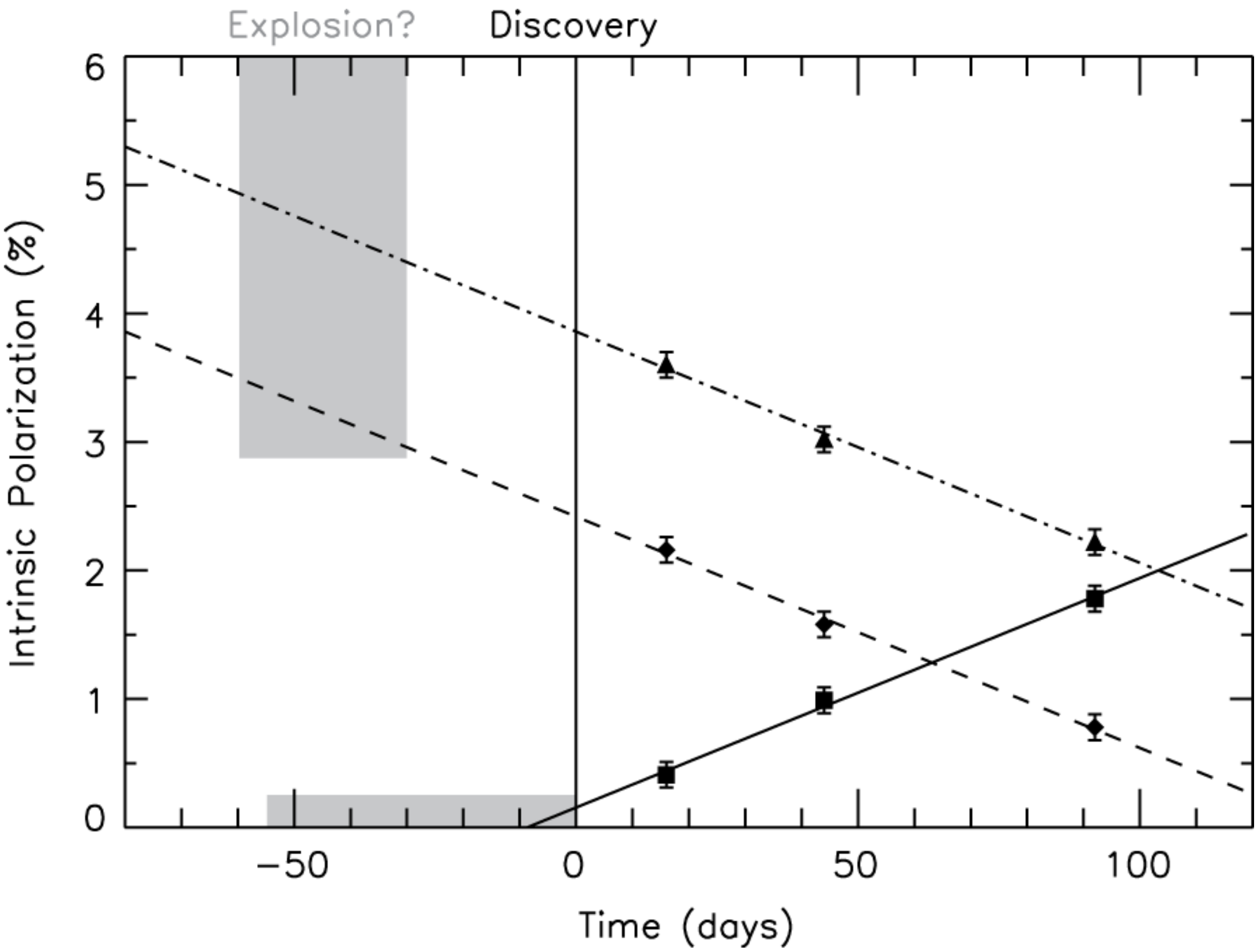

FIG. 11—Intrinsic continuum polarization of SN 1997eg as a function of time for each of the three representative ISP estimates described in § 4.3 and the Appendix: Cases A (*squares*), B (*diamonds*), and C (*triangles*). Error bars indicate instrumental errors of 0.1%. The three sloped lines are linear least-squares fits to the data points and have the equations $p_A = 0.018t + 0.16$ (*solid line*), $p_B = -0.018t + 2.42$ (*dashed line*), and $p_C = -0.018t + 3.86$ (*dot-dashed line*). The vertical line at $t = 0$ represents the supernova's discovery; the gray boxes represent estimated explosion dates and corresponding intrinsic polarization values for our ISP estimates (see Appendix for discussion of these estimates).

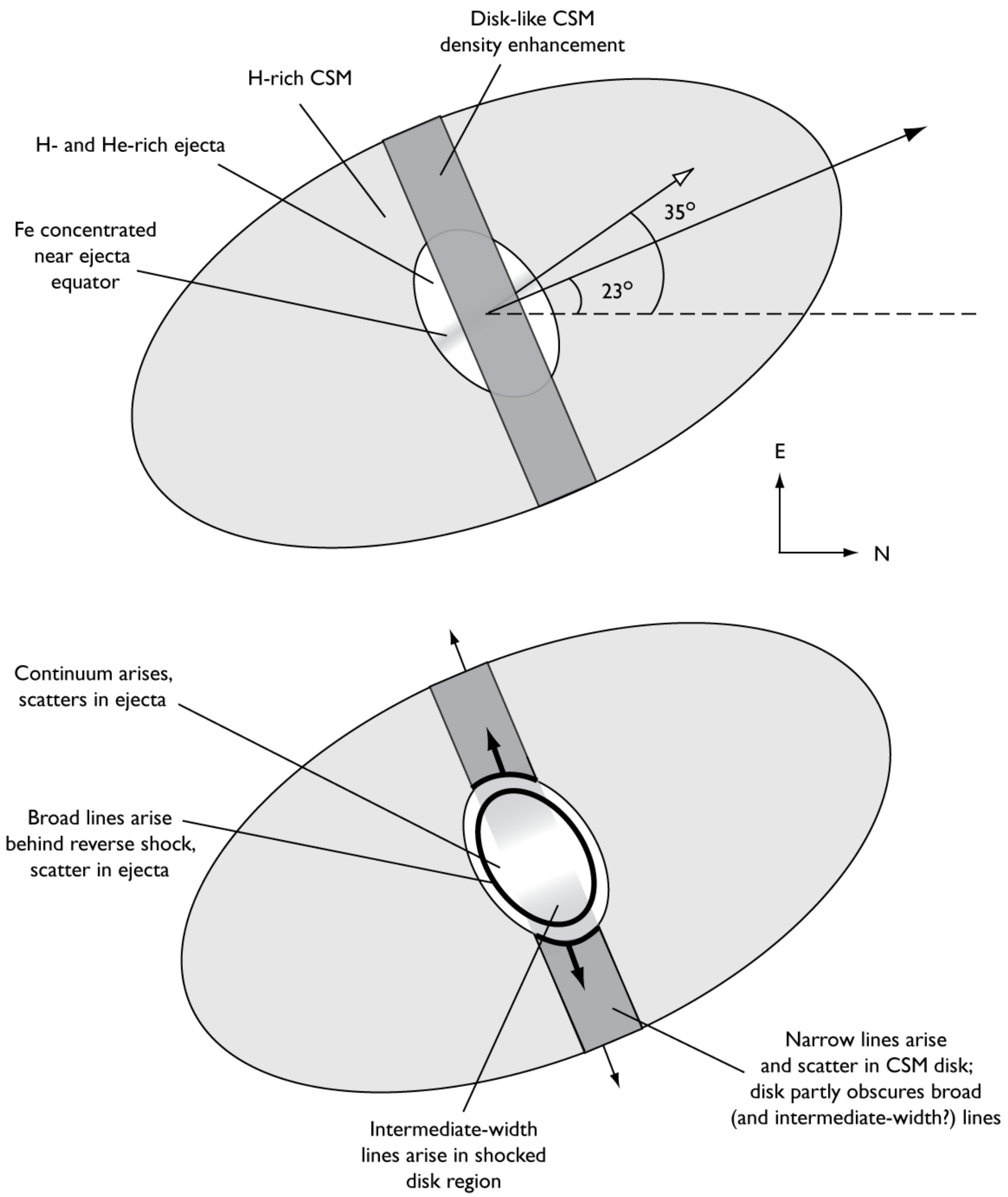

FIG. 12—Schematic representation of our model for SN 1997eg. The top sketch approximates our view from Earth; the arrows represent the orientations of the minor axis of the ejecta (35°) and the major axis of the circumstellar material (23°). The bottom sketch shows the components of the model in cross-section, with thick lines representing the forward and reverse shocks. For illustrative purposes only, we depict the CSM equatorial enhancement in each sketch as inclined at exactly 90° to our line of sight. We have also assumed Case A for the ISP contribution (§ 4.3; Appendix) and taken angle values less than 90° for simplicity. Relative sizes are not to scale, but relative orientations are correct with respect to the equatorial coordinate system (represented by the N–E compass).

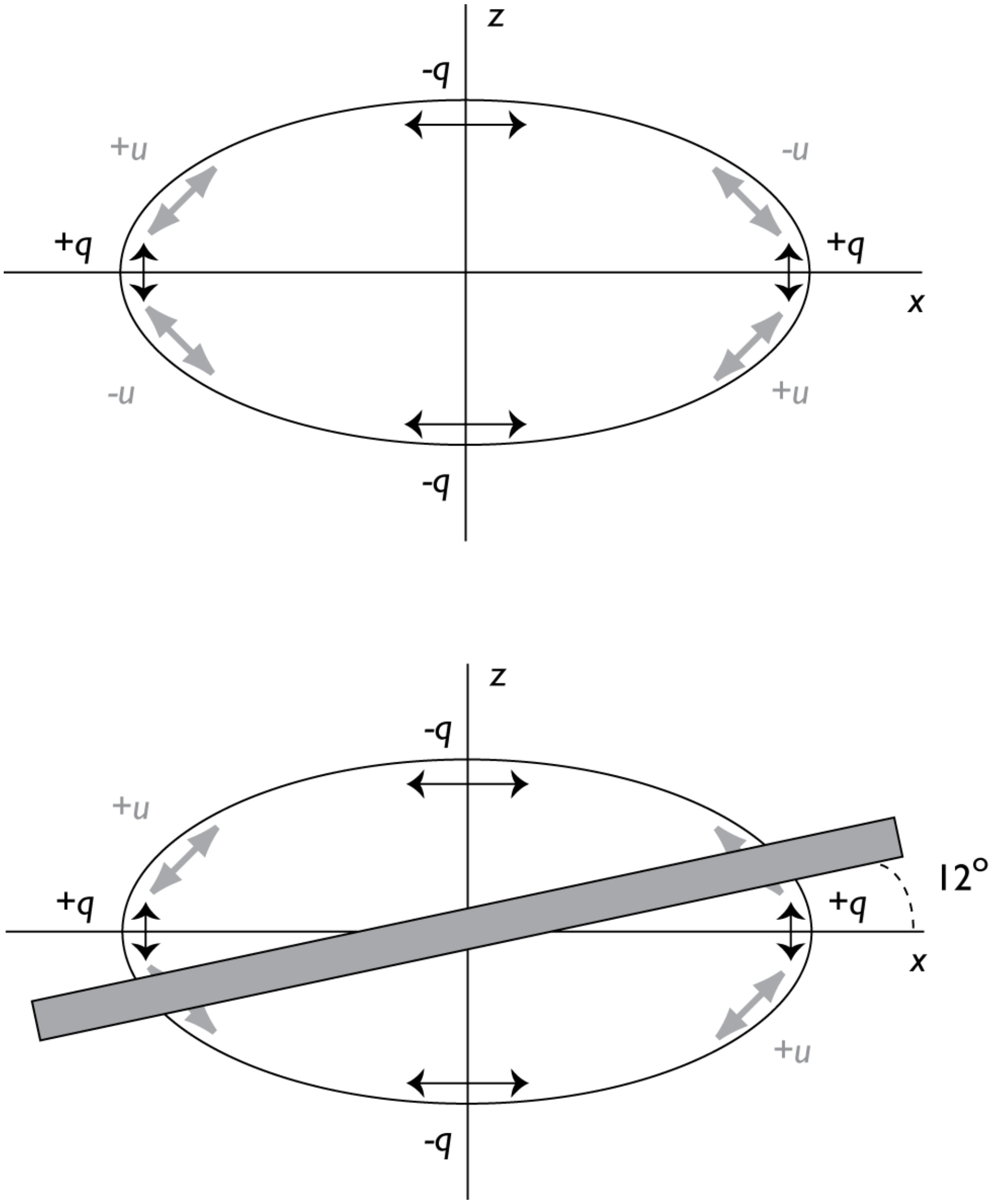

FIG. 13—Illustration of the intrinsic polarimetric properties of the model in Fig. 12 in its axial frame. Polarization in Stokes *q* (0°, ±90°) is shown as thin black arrows; polarization in Stokes *u* (±45°) is shown as thick gray arrows. Because we assume the ejecta to be ellipsoidal with axis ratio 0.5, the location of maximum +*u* polarization occurs at 14° from the major axis, not 45° as in the circular case. If the circumstellar disk is oriented at 12°, it blocks mainly +*u* polarization.